\documentclass[english,aps,prl,twocolumn,amsmath,amssymb,showpacs,superscriptaddress,notitlepage,longbibliography]{revtex4-2}
\usepackage{graphicx}
\usepackage{dcolumn}
\usepackage{bm}
\usepackage[usenames,dvipsnames]{color}
\usepackage[most]{tcolorbox}
\usepackage{multirow}
\usepackage{gensymb}
\usepackage{amssymb}
\usepackage{algorithm}
\usepackage{algorithmic}
\usepackage[normalem]{ulem}
\usepackage{CJK}
\usepackage{comment}
\usepackage{amsfonts}
\usepackage[colorlinks, linkcolor=blue,anchorcolor=blue,citecolor=blue,urlcolor=blue]{hyperref}
\usepackage{amssymb}
\usepackage{pifont}
\usepackage{physics}
\usepackage{amsmath}
\usepackage{natbib}
\usepackage{xcolor}
\usepackage{pdfcomment}

\usepackage{color,soul}

\begin{document}

\title{Quantum Well Resonant Tunneling Diode Probe of Correlated States in Twisted Bilayer MoS$_2$}

\author{Chengjie Zhou}
\affiliation{Physics Department and State Key Laboratory of Optical Quantum Materials, the Hong Kong University of Science and Technology, Clear Water Bay, Kowloon, Hong Kong SAR, China}
   
\author{Jin-Xin Hu}\thanks{jhuphy@ust.hk}
\affiliation{Physics Department, the Hong Kong University of Science and Technology, Clear Water Bay, Kowloon, Hong Kong SAR, China}

\author{Sihong Xu}

\author{Zijing Jin}
\affiliation{Physics Department and State Key Laboratory of Optical Quantum Materials, the Hong Kong University of Science and Technology, Clear Water Bay, Kowloon, Hong Kong SAR, China}

\author{Hui Li}
\affiliation{National Key Laboratory of Optoelectronic Information Acquisition and Protection Technology, Leibniz International Joint Research Center of Materials Sciences of Anhui Province, Institute of Physical Science and Information Technology, Anhui University, Anhui, China}

\author{Kam Tuen Law}
\affiliation{Physics Department, the Hong Kong University of Science and Technology, Clear Water Bay, Kowloon, Hong Kong SAR, China}

\author{Jiannong Wang}\thanks{phjwang@ust.hk}
\affiliation{Physics Department and State Key Laboratory of Optical Quantum Materials, the Hong Kong University of Science and Technology, Clear Water Bay, Kowloon, Hong Kong SAR, China}

\begin{abstract}
Moir\'{e} superlattices formed in transition metal dichalcogenides (TMDs) offer a versatile platform for exploring emergent quantum phases arising from strong electronic correlations. In this work, we develop a new experimental platform, the quantum well resonant tunneling diode (QWRTD), to probe the electronic landscape of $\approx 56.1^\circ$ twisted bilayer MoS$_2$ (tMoS$_2$). By measuring the differential conductance ($dI/dV_{\text{Probe}}$) as a function of filling factor $\nu$ and displacement field $D$, we observe a robust integer insulating feature at $\nu = 1$ that persists across the measured displacement field range. Furthermore, we identify a displacement-field-induced fractional insulating feature near $\nu = 3/4$ for $D < -75$~mV/nm. Its temperature dependence is consistent with a generalized Wigner crystal interpretation, although the microscopic order remains to be established. These features are also observed in another device with a similar twist angle ($\approx 56.4^\circ$). Our results provide evidence of correlated states in near-AB-stacked tMoS$_2$ and establish QWRTD as a tool for investigating strong correlations in van der Waals heterostructures.
\end{abstract}

\maketitle

\emph{Introduction.}---Moir\'{e} superlattices formed by stacking two-dimensional van der Waals materials with a small rotational or lattice mismatch have emerged as an exceptional platform for exploring strongly correlated physics\cite{bistritzerMoireBandsTwisted2011,caoUnconventionalSuperconductivityMagicangle2018,xuCorrelatedInsulatingStates2020,zengThermodynamicEvidenceFractional2023,caiSignaturesFractionalQuantum2023,parkObservationFractionallyQuantized2023}. In semiconductor transition metal dichalcogenides (TMDs), the interference pattern between neighboring layers generates a periodic moir\'{e} potential that quenches the kinetic energy of charge carriers, confining them into flat mini-bands\cite{wuHubbardModelPhysics2018,naikUltraflatbandsShearSolitons2018}. To date, most experimental studies on twisted TMD homobilayers and heterobilayers have focused on the hole-doped regime, revealing a rich spectrum of exotic states, including Mott insulators, generalized Wigner crystals, quantum anomalous/spin Hall states, and superconductivity\cite{reganMottGeneralizedWigner2020,xuCorrelatedInsulatingStates2020,liImagingTwodimensionalGeneralized2021,liMappingChargeExcitations2024,liWignerMolecularCrystals2024,caiSignaturesFractionalQuantum2023a,zengThermodynamicEvidenceFractional2023a,parkObservationFractionallyQuantized2023,xuObservationIntegerFractional2023,fouttyMappingTwisttunedMultiband2024,kangEvidenceFractionalQuantum2024,xiaSuperconductivityTwistedBilayer2025,guoSuperconductivity50degTwisted2025}.

\begin{figure}[t]
  \centering
  \includegraphics[width=1\linewidth]{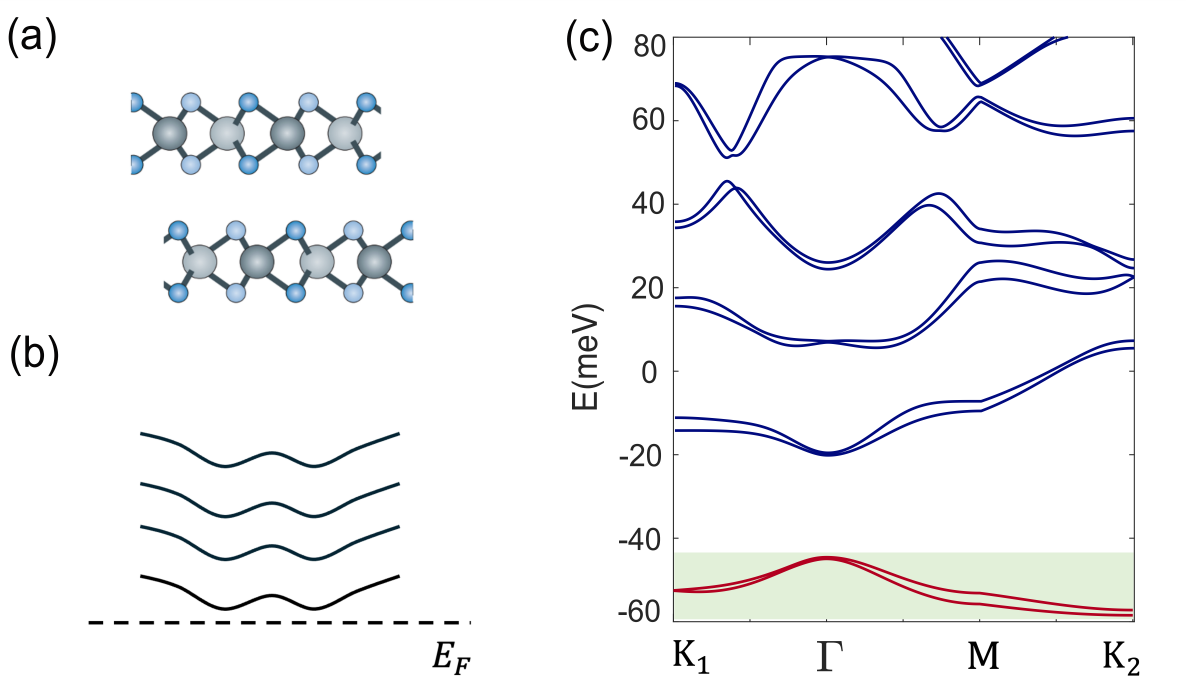}
  \caption{(a) Schematic picture of the crystal structure of AB-stacked tMoS$_2$ with 2H-type stacking order. (b) The moir\'{e} bands have nearly four-fold degeneracy including spin and valley. (c) Representative continuum-model band structure ($K$ valley) calculated for 57$^\circ$-tMoS$_2$ at displacement field $D=0.1$\text{ V/nm}; the measured device-1 angle is estimated as $56.1^\circ$.}\label{Figure1}
\end{figure}

\begin{figure*}[t]
  \centering
  \includegraphics[width=1\linewidth]{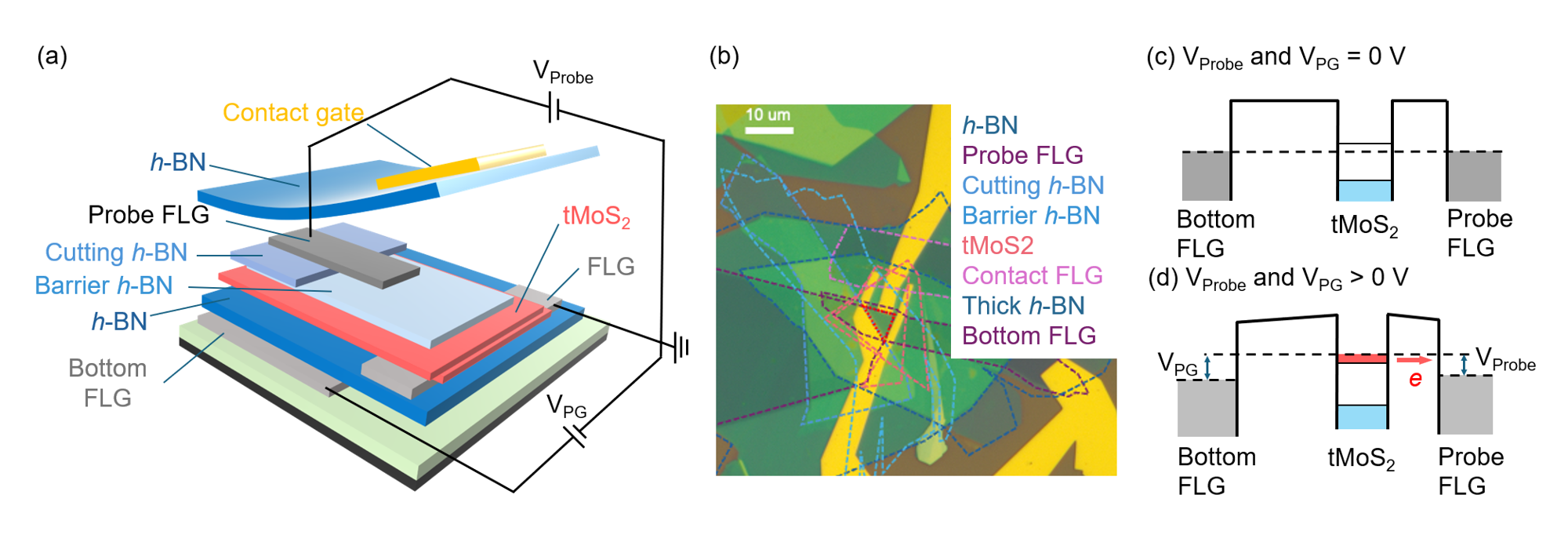}
  \caption{(a) Schematic structure of a QWRTD (contact gate/top $h$-BN/Probe FLG/cutting $h$-BN/barrier $h$-BN/tMoS$_2$/contact FLG/bottom $h$-BN/bottom FLG). The probe bias $V_{\text{Probe}}$ is applied between the Probe FLG and tMoS$_2$ (via contact FLG), while the probe-gate voltage $V_{\text{PG}}$ is applied between the bottom FLG and tMoS$_2$, together enabling joint control of the tunneling bias and the local displacement field/doping. (b) Optical micrograph of the assembled QWRTD device. Distinct constituent layers are outlined by dashed lines of different colors, with the active tunneling region enclosed by the red dashed contour. (c-d) Schematic band alignment of a QWRTD when $V_{\text{Probe}}$ and $V_{\text{PG}} = 0\text{ V}$ (c), and $V_{\text{Probe}}$ and $V_{\text{PG}} > 0\text{ V}$ (d).}\label{Figure2}
\end{figure*}

Pioneering research on twisted TMDs focused predominantly on the valence band owing to its simple electronic structure, in which the $K$ and $K'$ valleys' localization driven by in-plane $d_{x^2-y^2}$ and $d_{xy}$ orbitals and giant Ising spin-orbit coupling of hundreds of meV yields a single-band triangular Hubbard model~\cite{wuHubbardModelPhysics2018}. By contrast, the conduction band exhibits enhanced interlayer hybridization from dominant out-of-plane $d_{z^2}$ orbitals, multi-pocket quantum interference from nearby six-fold $Q$ ($\Lambda$) valleys, and significantly weaker spin-orbit coupling. This reduced spin-valley locking renders electron-doped moir\'{e} superlattices highly field-tunable, providing a versatile platform for field-induced topological band inversions and multi-flavor electronic phases. Near AB-stacking ($\approx 56.1^\circ$) twisted bilayer MoS$_2$ (tMoS$_2$) offers a promising platform to explore this electron-doped regime, where narrow moir\'e minibands and strong Coulomb interactions are expected to yield rich correlated electron dynamics \cite{wuGiantCorrelatedGap2023}. Unlike AA-stacked TMD bilayers, 2H-type AB stacking breaks in-plane inversion symmetry [Fig.~\ref{Figure1}(a)], giving rise to distinct layer-polarized moir\'e potentials. Due to the combined spin and valley degrees of freedom, the low-energy moir\'e minibands display a nearly four-fold degeneracy [Fig.~\ref{Figure1}(b)]. A representative continuum-model calculation at $57^\circ$ [see Supplemental Material~\cite{SupplementalMaterial}, Sec.~S1] indicates that the lowest conduction moir\'e miniband is well isolated from higher-lying bands under a displacement field $D = 0.1\text{ V/nm}$ [Fig.~\ref{Figure1}(c)], exhibiting a narrow bandwidth $W \approx 15\text{ meV}$ at the $K$ valley, weakly split by spin-orbit coupling. Comparison with the estimated on-site Coulomb interaction $U \approx 25\text{ meV}$ yields a correlation ratio $U/W \gtrsim 1$ for the representative model; a quantitative calculation at the measured device angle remains to be provided.

However, experimental mapping of this electron-doped flat-band regime remains severely limited. Conventional macroscopic planar transport measurements fail to resolve these flat-band states, hindered by the high Schottky contact barriers at the TMD conduction band edge\cite{wangVanWaalsContacts2019,wangPtypeElectricalContacts2022,parkObservationFractionallyQuantized2023,wangCriticalChallengesDevelopment2024,packChargetransferContactsMeasurement2024} and the inevitable spatial averaging across macroscopic channels due to non-uniform distributions of twist angle and electric displacement field $D$ in large-scale devices \cite{uriSuperconductivityStrongInteractions2023,yooAtomicElectronicReconstruction2019a,wongCascadeElectronicTransitions2020a,kazmierczakStrainFieldsTwisted2021}. While local scanning probes, such as scanning tunneling spectroscopy (STS) and scanning single-electron transistors (SET), effectively mitigate spatial inhomogeneity by isolating local domains\cite{wongCascadeElectronicTransitions2020,xieFractionalChernInsulators2021,fouttyDatasetMappingTwisttuned2024,liMappingChargeExcitations2024}, their open-geometry configuration intrinsically precludes the top gate integration required for independent, quantitative displacement-field ($D$) control. A technique capable of unifying local spatial resolution with full dual-gate field tunability and efficient electron injection has thus remained elusive.

In this Letter, we introduce a quantum well resonant tunneling diode (QWRTD) architecture\cite{luryiQuantumCapacitanceDevices1988,jungDirectProbingElectronic2017} that enables localized tunneling spectroscopy of near AB-stacking ($\approx 56.1^\circ$) tMoS$_2$ moir\'e systems. This platform senses correlated features via vertical tunneling while retaining dual-gate electrostatics. In the electron-doped regime at zero magnetic field, we observe a robust integer insulating feature at $\nu = 1$, consistent with a Mott state, and an emergent displacement-field-induced fractional insulating feature near $\nu = 3/4$. The temperature dependence of the latter is consistent with a generalized Wigner crystal interpretation, establishing the QWRTD as a probe for multi-flavor moir\'e physics.

\begin{figure*}[t]
  \centering
  \includegraphics[width=1\linewidth]{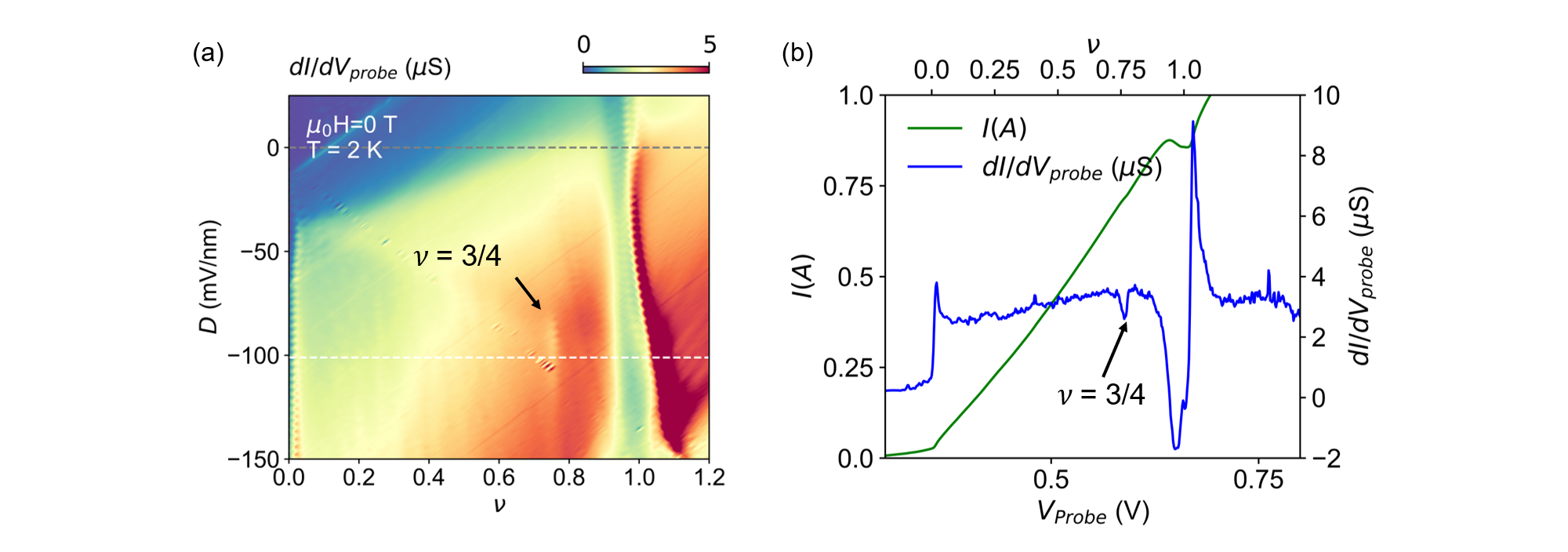} 
  \caption{(a) Measured differential conductance $dI/dV_{\text{Probe}}$ map versus apparent filling factor $\nu$ and effective vertical field $D$ at $T = 2\text{ K}$ and zero magnetic field. An integer conductance minimum persists near $\nu = 1$, while a fractional minimum emerges near $\nu = 3/4$ for $D \lesssim -75\text{ mV/nm}$ (black arrow). (b) Measured $I$-$V_{\text{Probe}}$ curve (green) and corresponding $dI/dV_{\text{Probe}}$ curve (blue) along the white approximately constant-$D$ trajectory in (a), reported as $D\approx-101.6$~mV/nm.}\label{Figure3}
\end{figure*}

\emph{Device architecture and electrostatic mapping.}---The QWRTD architecture and operational mechanism are illustrated in Fig.~\ref{Figure2}. The heterostructure comprises an 8-layer vertical stack (top h-BN/Probe few-layer graphene (FLG)/cutting h-BN/barrier h-BN/tMoS$_2$/contact FLG/bottom h-BN/bottom FLG) [Fig.~\ref{Figure2}(a)], where the key to non-invasive local probing lies in suppressing parasitic transport pathways. Specifically, a patterned cutting $h$-BN layer is designed to suppress edge and monolayer tunneling, concentrating the measured signal from the pristine moir\'e channel [see Supplemental Material~\cite{SupplementalMaterial}, Sec.~S2]. Fig.~\ref{Figure2}(b) shows the optical micrograph of the assembled functional QWRTD, where constituent 2D layers are outlined by dashed contours. Consistent with our design [Fig.~\ref{Figure2}(a)], the intended active tunneling region is bounded by the patterned cutting $h$-BN layer, highlighted by the red dashed boundary in Fig.~\ref{Figure2}(b). As depicted in the schematic band alignments [Figs.~\ref{Figure2}(c) and \ref{Figure2}(d)], applying a probe bias $V_{\text{Probe}} > 0\text{ V}$ shifts the probe Fermi level relative to the tMoS$_2$ conduction band, driving a vertical tunneling current $I$ from tMoS$_2$ to probe-FLG through the thin barrier $h$-BN layer. Concurrently, the voltage $V_{\text{PG}}$, applied to the bottom FLG relative to tMoS$_2$, capacitively couples to the tMoS$_2$ channel in conjunction with $V_{\text{Probe}}$: their sum (weighted by the respective gate capacitances) sets the local carrier density $n$, while their difference sets the displacement field $D$, enabling combined control of doping and tunneling bias. The measured current--voltage curves are numerically differentiated to obtain the local differential conductance ($dI/dV_{\text{Probe}}$).

Continuous control over both the filling factor $\nu$ and the effective vertical field $D$ inside the tMoS$_2$ channel is achieved through dual-gate electrostatics. By solving the electrostatics of the probe-FLG/barrier-$h$-BN/tMoS$_2$/bottom-FLG junction, the net carrier density (proportional to $\nu$) and $D$ are governed by linear combinations of $V_{\text{Probe}}$ and $V_{\text{PG}}$ [see Supplemental Material~\cite{SupplementalMaterial}, Sec.~S3]. The AFM barrier thickness is $d_0\approx2$~nm in both devices; we use a thin-$h$-BN relative permittivity of 3.3 and infer the bottom-to-top capacitance ratio from the gate-map slopes. Specifically, trajectories along $V_{\text{PG}} = -k V_{\text{Probe}} + \text{constant}$ (where $k>0$ is the magnitude of the negative slope parallel to the conduction-band edge or the $\nu=1$ feature) preserve a constant net charge (and thus constant $\nu$), whereas trajectories along $V_{\text{PG}} = k V_{\text{Probe}} + \text{constant}$ maintain constant $D$. We use $k=13.3$ and 6.5 for devices 1 and 2, respectively. This electrostatic transformation allows us to display raw $dI/dV_{\text{Probe}}(V_{\text{Probe}}, V_{\text{PG}})$ signals as $dI/dV_{\text{Probe}}(\nu, D)$ maps. Because $V_{\text{Probe}}$ also sets the tunneling bias, each trajectory changes the probed energy along with either $D$ or $\nu$.

\emph{Integer and fractional correlated insulating features.}---Fig.~\ref{Figure3}(a) presents the measured local differential conductance $dI/dV_{\text{Probe}}$ spectrum as a function of filling factor $\nu$ and effective vertical field $D$ at zero magnetic field and $T = 2\text{ K}$. We use the positions of conductance minima as a consistent operational reference for insulating features. Two prominent minima emerge near $\nu = 1$ and $\nu = 3/4$. Because the probe bias also changes the channel density, these minima need not coincide exactly with the equilibrium centers of the insulating states; the possible offset is discussed in Supplemental Material~\cite{SupplementalMaterial}, Sec.~S4. The integer feature ($\nu = 1$) persists across the measured range of $D$ and is consistent with the previously reported Mott insulating state \cite{wuGiantCorrelatedGap2023}. The fractional feature near $\nu = 3/4$ develops as $D$ is tuned below approximately $-75\text{ mV/nm}$, as indicated by the black arrow in Fig.~\ref{Figure3}(a). Fig.~\ref{Figure3}(b) shows the measured $I$-$V_{\text{Probe}}$ curve (green) and corresponding $dI/dV_{\text{Probe}}$ curve (blue) along the white approximately constant-field trajectory in Fig.~\ref{Figure3}(a), reported as $D\approx-101.6$~mV/nm.

\begin{figure*}[t]
  \centering
  \includegraphics[width=0.8\linewidth]{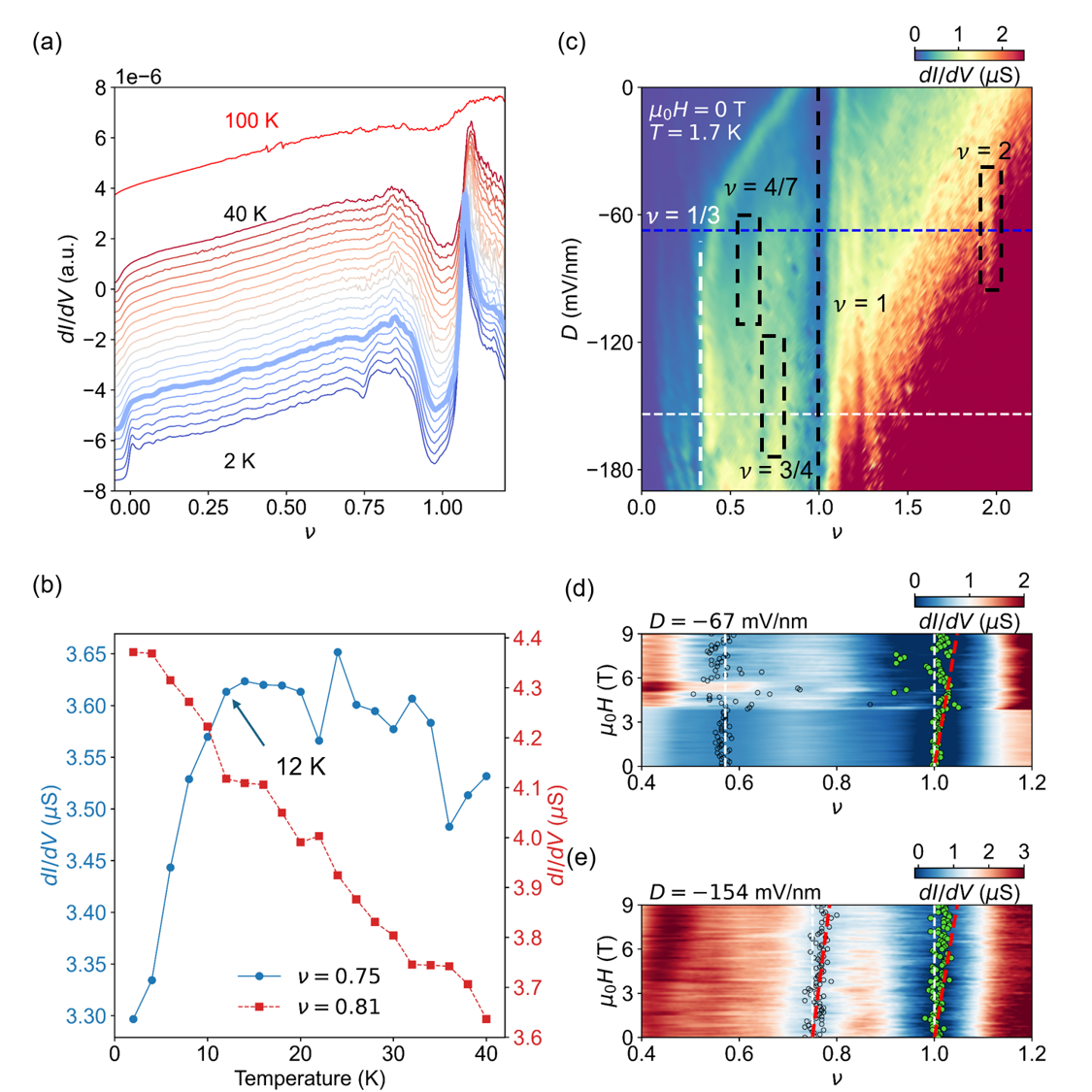}
  \caption{(a) $dI/dV_{\text{Probe}}$ versus filling factor $\nu$ from $2$ to $100$~K along a fixed-$D$ sweep. Curves are vertically offset; the $12$~K curve, above which the $\nu = 3/4$ minimum is no longer resolved, is highlighted in bold. (b) Temperature dependence of $dI/dV_{\text{Probe}}$ at $\nu = 0.75$ (blue) and $\nu = 0.81$ (red). (c) $dI/dV_{\text{Probe}}$ versus $\nu$ and $D$ for a second $\approx 56.4^\circ$ tMoS$_2$ device at $1.7$~K and $0$~T. Vertical marks highlight features at $\nu = 1/3, 4/7, 3/4, 1,$ and $2$. (d-e) $dI/dV_{\text{Probe}}$ versus $\nu$ and magnetic field at $D \approx -67$ and $-154$~mV/nm, respectively. Red dashed lines show simulated St\v{r}eda trajectories, and white dashed lines show zero-slope references.}
  \label{Figure4}
\end{figure*}

Fig.~\ref{Figure4}(a) displays the temperature-dependent $dI/dV_{\text{Probe}}$ signal versus filling factor $\nu$ for temperatures ranging from $2$~K to $100$~K (with $2$~K increments below $40$~K) along a separate fixed-$D$ sweep at zero magnetic field. The $\nu = 1$ minimum remains visible up to $100$~K, whereas the $\nu = 3/4$ minimum is no longer resolved above $12$~K (highlighted by the bold curve). Fig.~\ref{Figure4}(b) plots $dI/dV_{\text{Probe}}$ at $\nu = 0.75$ (blue) and $\nu = 0.81$ (red). At $\nu = 0.81$, the conductance decreases monotonically with temperature, whereas at $\nu = 0.75$ it initially rises before declining above $12$~K. The different temperature scales are consistent with a robust integer correlated state and a more fragile fractional state. The latter behavior is consistent with a generalized Wigner crystal interpretation~\cite{reganMottGeneralizedWigner2020,xuCorrelatedInsulatingStates2020,liImagingTwodimensionalGeneralized2021}, but temperature-dependent tunneling conductance alone does not establish charge order.

Conductance minima near $\nu = 1$ and $\nu = 3/4$ also appear in a second QWRTD device with a $\approx 56.4^\circ$ tMoS$_2$ moir\'e system [Fig.~\ref{Figure4}(c)]. At smaller $|D|$, additional minima appear near $\nu = 4/7$ and $\nu = 2$, together with a local conductance enhancement around $\nu = 1/3$. The $4/7$ feature is no longer resolved at larger $|D|$, while the $3/4$ feature persists. This reproducibility supports an association of the fractional features with the electron-doped moir\'e system, subject to the filling calibration and tunneling-bias considerations described in Supplemental Material~\cite{SupplementalMaterial}, Secs.~S3 and S4.

Figs.~\ref{Figure4}(d) and (e) map $dI/dV_{\text{Probe}}$ versus filling factor $\nu$ and magnetic field at $D \approx -67$ and $-154$~mV/nm, respectively. At $D \approx -67$~mV/nm, minima occur near $\nu = 4/7$ and $\nu = 1$. At $D \approx -154$~mV/nm, the $4/7$ feature is no longer resolved, while a fractional feature near $3/4$ appears alongside the integer feature. Red dashed lines show simulated St\v{r}eda trajectories; white dashed lines have zero slope. The $\nu = 1$ feature shifts with field but follows neither reference trajectory. The $4/7$ feature at $D \approx -67$~mV/nm remains close to the zero-slope reference, whereas the $3/4$ feature at $D \approx -154$~mV/nm appears closer to a St\v{r}eda trajectory. Field-dependent response of the FLG probe or contacts could shift apparent feature positions, but their contribution has not yet been separated from the tMoS$_2$ response.

The interplay between flat bands and long-range Coulomb interactions can stabilize charge-ordered states at commensurate fractional fillings. In a generalized Wigner crystal (GWC), carriers occupy an enlarged real-space unit cell to reduce Coulomb repulsion. Other correlated phases can also occur at fractional filling, so a conductance minimum alone does not establish broken translational symmetry.

At $\nu=3/4$, a candidate GWC configuration would contain three occupied sites per four moir\'{e} cells. Establishing this spatial order requires evidence beyond the present tunneling and temperature measurements. The fractional feature therefore motivates further tests of charge ordering and its competition with other correlated phases.

\emph{Conclusion.}---\label{sec13}In summary, we have demonstrated a gate-tunable QWRTD as a localized probe of correlated features in near-AB-stacked twisted bilayer MoS$_2$ (tMoS$_2$). By mapping $dI/dV_{\text{Probe}}$ as a function of filling factor $\nu$ and effective vertical field $D$, we observed, in an $\approx56.1^\circ$ device, a robust integer feature at $\nu = 1$ across the measured $D$ range and a fractional feature near $\nu = 3/4$ emerging for $D \lesssim -75\text{ mV/nm}$. Their temperature dependence is consistent with a Mott interpretation for the integer feature and a possible generalized Wigner crystal interpretation for the fractional feature; microscopic order remains to be established. Similar features occur in a second device with an $\approx56.4^\circ$ twist angle, although the field ranges of the fractional features are less well resolved there. These observations extend tunneling studies of fractional correlated features to electron-doped TMD moir\'e superlattices.

\begin{acknowledgments}
This work was supported by the Research Grants Council of the Hong Kong SAR under Grant Nos.\ 16302023, C6053-23G, and AoE/P-604/25R. We acknowledge the technical support from the HKUST Nano Fabrication Facility and Materials Characterization and Preparation Facility.
\end{acknowledgments}

\nocite{wangOneDimensionalElectricalContact2013,lauReproducibilityFabricationPhysics2022,caiSignaturesFractionalQuantum2023,laturiaDielectricPropertiesHexagonal2018,pierretDielectricPermittivityConductivity2022}
\bibliographystyle{apsrev4-1} 
\bibliography{references}

\begin{thebibliography}{38}%
\makeatletter
\providecommand \@ifxundefined [1]{%
 \@ifx{#1\undefined}
}%
\providecommand \@ifnum [1]{%
 \ifnum #1\expandafter \@firstoftwo
 \else \expandafter \@secondoftwo
 \fi
}%
\providecommand \@ifx [1]{%
 \ifx #1\expandafter \@firstoftwo
 \else \expandafter \@secondoftwo
 \fi
}%
\providecommand \natexlab [1]{#1}%
\providecommand \enquote  [1]{``#1''}%
\providecommand \bibnamefont  [1]{#1}%
\providecommand \bibfnamefont [1]{#1}%
\providecommand \citenamefont [1]{#1}%
\providecommand \href@noop [0]{\@secondoftwo}%
\providecommand \href [0]{\begingroup \@sanitize@url \@href}%
\providecommand \@href[1]{\@@startlink{#1}\@@href}%
\providecommand \@@href[1]{\endgroup#1\@@endlink}%
\providecommand \@sanitize@url [0]{\catcode `\\12\catcode `\$12\catcode
  `\&12\catcode `\#12\catcode `\^12\catcode `\_12\catcode `\%12\relax}%
\providecommand \@@startlink[1]{}%
\providecommand \@@endlink[0]{}%
\providecommand \url  [0]{\begingroup\@sanitize@url \@url }%
\providecommand \@url [1]{\endgroup\@href {#1}{\urlprefix }}%
\providecommand \urlprefix  [0]{URL }%
\providecommand \Eprint [0]{\href }%
\providecommand \doibase [0]{http://dx.doi.org/}%
\providecommand \selectlanguage [0]{\@gobble}%
\providecommand \bibinfo  [0]{\@secondoftwo}%
\providecommand \bibfield  [0]{\@secondoftwo}%
\providecommand \translation [1]{[#1]}%
\providecommand \BibitemOpen [0]{}%
\providecommand \bibitemStop [0]{}%
\providecommand \bibitemNoStop [0]{.\EOS\space}%
\providecommand \EOS [0]{\spacefactor3000\relax}%
\providecommand \BibitemShut  [1]{\csname bibitem#1\endcsname}%
\let\auto@bib@innerbib\@empty
\bibitem [{\citenamefont {Bistritzer}\ and\ \citenamefont
  {MacDonald}(2011)}]{bistritzerMoireBandsTwisted2011}%
  \BibitemOpen
  \bibfield  {author} {\bibinfo {author} {\bibfnamefont {R.}~\bibnamefont
  {Bistritzer}}\ and\ \bibinfo {author} {\bibfnamefont {A.~H.}\ \bibnamefont
  {MacDonald}},\ }\href {\doibase 10.1073/pnas.1108174108} {\bibfield
  {journal} {\bibinfo  {journal} {Proceedings of the National Academy of
  Sciences}\ }\textbf {\bibinfo {volume} {108}},\ \bibinfo {pages} {12233}
  (\bibinfo {year} {2011})}\BibitemShut {NoStop}%
\bibitem [{\citenamefont {Cao}\ \emph {et~al.}(2018)\citenamefont {Cao},
  \citenamefont {Fatemi}, \citenamefont {Fang}, \citenamefont {Watanabe},
  \citenamefont {Taniguchi}, \citenamefont {Kaxiras},\ and\ \citenamefont
  {Jarillo-Herrero}}]{caoUnconventionalSuperconductivityMagicangle2018}%
  \BibitemOpen
  \bibfield  {author} {\bibinfo {author} {\bibfnamefont {Y.}~\bibnamefont
  {Cao}}, \bibinfo {author} {\bibfnamefont {V.}~\bibnamefont {Fatemi}},
  \bibinfo {author} {\bibfnamefont {S.}~\bibnamefont {Fang}}, \bibinfo {author}
  {\bibfnamefont {K.}~\bibnamefont {Watanabe}}, \bibinfo {author}
  {\bibfnamefont {T.}~\bibnamefont {Taniguchi}}, \bibinfo {author}
  {\bibfnamefont {E.}~\bibnamefont {Kaxiras}}, \ and\ \bibinfo {author}
  {\bibfnamefont {P.}~\bibnamefont {Jarillo-Herrero}},\ }\href {\doibase
  10.1038/nature26160} {\bibfield  {journal} {\bibinfo  {journal} {Nature}\
  }\textbf {\bibinfo {volume} {556}},\ \bibinfo {pages} {43} (\bibinfo {year}
  {2018})}\BibitemShut {NoStop}%
\bibitem [{\citenamefont {Xu}\ \emph {et~al.}(2020)\citenamefont {Xu},
  \citenamefont {Liu}, \citenamefont {Rhodes}, \citenamefont {Watanabe},
  \citenamefont {Taniguchi}, \citenamefont {Hone}, \citenamefont {Elser},
  \citenamefont {Mak},\ and\ \citenamefont
  {Shan}}]{xuCorrelatedInsulatingStates2020}%
  \BibitemOpen
  \bibfield  {author} {\bibinfo {author} {\bibfnamefont {Y.}~\bibnamefont
  {Xu}}, \bibinfo {author} {\bibfnamefont {S.}~\bibnamefont {Liu}}, \bibinfo
  {author} {\bibfnamefont {D.~A.}\ \bibnamefont {Rhodes}}, \bibinfo {author}
  {\bibfnamefont {K.}~\bibnamefont {Watanabe}}, \bibinfo {author}
  {\bibfnamefont {T.}~\bibnamefont {Taniguchi}}, \bibinfo {author}
  {\bibfnamefont {J.}~\bibnamefont {Hone}}, \bibinfo {author} {\bibfnamefont
  {V.}~\bibnamefont {Elser}}, \bibinfo {author} {\bibfnamefont {K.~F.}\
  \bibnamefont {Mak}}, \ and\ \bibinfo {author} {\bibfnamefont
  {J.}~\bibnamefont {Shan}},\ }\href {\doibase 10.1038/s41586-020-2868-6}
  {\bibfield  {journal} {\bibinfo  {journal} {Nature}\ }\textbf {\bibinfo
  {volume} {587}},\ \bibinfo {pages} {214} (\bibinfo {year}
  {2020})}\BibitemShut {NoStop}%
\bibitem [{\citenamefont {Zeng}\ \emph
  {et~al.}(2023{\natexlab{a}})\citenamefont {Zeng}, \citenamefont {Xia},
  \citenamefont {Kang}, \citenamefont {Zhu}, \citenamefont {Knüppel},
  \citenamefont {Vaswani}, \citenamefont {Watanabe}, \citenamefont {Taniguchi},
  \citenamefont {Mak},\ and\ \citenamefont
  {Shan}}]{zengThermodynamicEvidenceFractional2023}%
  \BibitemOpen
  \bibfield  {author} {\bibinfo {author} {\bibfnamefont {Y.}~\bibnamefont
  {Zeng}}, \bibinfo {author} {\bibfnamefont {Z.}~\bibnamefont {Xia}}, \bibinfo
  {author} {\bibfnamefont {K.}~\bibnamefont {Kang}}, \bibinfo {author}
  {\bibfnamefont {J.}~\bibnamefont {Zhu}}, \bibinfo {author} {\bibfnamefont
  {P.}~\bibnamefont {Knüppel}}, \bibinfo {author} {\bibfnamefont
  {C.}~\bibnamefont {Vaswani}}, \bibinfo {author} {\bibfnamefont
  {K.}~\bibnamefont {Watanabe}}, \bibinfo {author} {\bibfnamefont
  {T.}~\bibnamefont {Taniguchi}}, \bibinfo {author} {\bibfnamefont {K.~F.}\
  \bibnamefont {Mak}}, \ and\ \bibinfo {author} {\bibfnamefont
  {J.}~\bibnamefont {Shan}},\ }\href {\doibase 10.1038/s41586-023-06452-3}
  {\bibfield  {journal} {\bibinfo  {journal} {Nature}\ }\textbf {\bibinfo
  {volume} {622}},\ \bibinfo {pages} {69} (\bibinfo {year}
  {2023}{\natexlab{a}})}\BibitemShut {NoStop}%
\bibitem [{\citenamefont {Cai}\ \emph {et~al.}(2023{\natexlab{a}})\citenamefont
  {Cai}, \citenamefont {Anderson}, \citenamefont {Wang}, \citenamefont {Zhang},
  \citenamefont {Liu}, \citenamefont {Holtzmann}, \citenamefont {Zhang},
  \citenamefont {Fan}, \citenamefont {Taniguchi}, \citenamefont {Watanabe},
  \citenamefont {Ran}, \citenamefont {Cao}, \citenamefont {Fu}, \citenamefont
  {Xiao}, \citenamefont {Yao},\ and\ \citenamefont
  {Xu}}]{caiSignaturesFractionalQuantum2023}%
  \BibitemOpen
  \bibfield  {author} {\bibinfo {author} {\bibfnamefont {J.}~\bibnamefont
  {Cai}}, \bibinfo {author} {\bibfnamefont {E.}~\bibnamefont {Anderson}},
  \bibinfo {author} {\bibfnamefont {C.}~\bibnamefont {Wang}}, \bibinfo {author}
  {\bibfnamefont {X.}~\bibnamefont {Zhang}}, \bibinfo {author} {\bibfnamefont
  {X.}~\bibnamefont {Liu}}, \bibinfo {author} {\bibfnamefont {W.}~\bibnamefont
  {Holtzmann}}, \bibinfo {author} {\bibfnamefont {Y.}~\bibnamefont {Zhang}},
  \bibinfo {author} {\bibfnamefont {F.}~\bibnamefont {Fan}}, \bibinfo {author}
  {\bibfnamefont {T.}~\bibnamefont {Taniguchi}}, \bibinfo {author}
  {\bibfnamefont {K.}~\bibnamefont {Watanabe}}, \bibinfo {author}
  {\bibfnamefont {Y.}~\bibnamefont {Ran}}, \bibinfo {author} {\bibfnamefont
  {T.}~\bibnamefont {Cao}}, \bibinfo {author} {\bibfnamefont {L.}~\bibnamefont
  {Fu}}, \bibinfo {author} {\bibfnamefont {D.}~\bibnamefont {Xiao}}, \bibinfo
  {author} {\bibfnamefont {W.}~\bibnamefont {Yao}}, \ and\ \bibinfo {author}
  {\bibfnamefont {X.}~\bibnamefont {Xu}},\ }\href {\doibase
  10.1038/s41586-023-06289-w} {\bibfield  {journal} {\bibinfo  {journal}
  {Nature}\ }\textbf {\bibinfo {volume} {622}},\ \bibinfo {pages} {63}
  (\bibinfo {year} {2023}{\natexlab{a}})}\BibitemShut {NoStop}%
\bibitem [{\citenamefont {Park}\ \emph {et~al.}(2023)\citenamefont {Park},
  \citenamefont {Cai}, \citenamefont {Anderson}, \citenamefont {Zhang},
  \citenamefont {Zhu}, \citenamefont {Liu}, \citenamefont {Wang}, \citenamefont
  {Holtzmann}, \citenamefont {Hu}, \citenamefont {Liu}, \citenamefont
  {Taniguchi}, \citenamefont {Watanabe}, \citenamefont {Chu}, \citenamefont
  {Cao}, \citenamefont {Fu}, \citenamefont {Yao}, \citenamefont {Chang},
  \citenamefont {Cobden}, \citenamefont {Xiao},\ and\ \citenamefont
  {Xu}}]{parkObservationFractionallyQuantized2023}%
  \BibitemOpen
  \bibfield  {author} {\bibinfo {author} {\bibfnamefont {H.}~\bibnamefont
  {Park}}, \bibinfo {author} {\bibfnamefont {J.}~\bibnamefont {Cai}}, \bibinfo
  {author} {\bibfnamefont {E.}~\bibnamefont {Anderson}}, \bibinfo {author}
  {\bibfnamefont {Y.}~\bibnamefont {Zhang}}, \bibinfo {author} {\bibfnamefont
  {J.}~\bibnamefont {Zhu}}, \bibinfo {author} {\bibfnamefont {X.}~\bibnamefont
  {Liu}}, \bibinfo {author} {\bibfnamefont {C.}~\bibnamefont {Wang}}, \bibinfo
  {author} {\bibfnamefont {W.}~\bibnamefont {Holtzmann}}, \bibinfo {author}
  {\bibfnamefont {C.}~\bibnamefont {Hu}}, \bibinfo {author} {\bibfnamefont
  {Z.}~\bibnamefont {Liu}}, \bibinfo {author} {\bibfnamefont {T.}~\bibnamefont
  {Taniguchi}}, \bibinfo {author} {\bibfnamefont {K.}~\bibnamefont {Watanabe}},
  \bibinfo {author} {\bibfnamefont {J.-h.}\ \bibnamefont {Chu}}, \bibinfo
  {author} {\bibfnamefont {T.}~\bibnamefont {Cao}}, \bibinfo {author}
  {\bibfnamefont {L.}~\bibnamefont {Fu}}, \bibinfo {author} {\bibfnamefont
  {W.}~\bibnamefont {Yao}}, \bibinfo {author} {\bibfnamefont {C.-Z.}\
  \bibnamefont {Chang}}, \bibinfo {author} {\bibfnamefont {D.}~\bibnamefont
  {Cobden}}, \bibinfo {author} {\bibfnamefont {D.}~\bibnamefont {Xiao}}, \ and\
  \bibinfo {author} {\bibfnamefont {X.}~\bibnamefont {Xu}},\ }\href {\doibase
  10.1038/s41586-023-06536-0} {\bibfield  {journal} {\bibinfo  {journal}
  {Nature}\ ,\ \bibinfo {pages} {1}} (\bibinfo {year} {2023})}\BibitemShut
  {NoStop}%
\bibitem [{\citenamefont {Wu}\ \emph {et~al.}(2018)\citenamefont {Wu},
  \citenamefont {Lovorn}, \citenamefont {Tutuc},\ and\ \citenamefont
  {MacDonald}}]{wuHubbardModelPhysics2018}%
  \BibitemOpen
  \bibfield  {author} {\bibinfo {author} {\bibfnamefont {F.}~\bibnamefont
  {Wu}}, \bibinfo {author} {\bibfnamefont {T.}~\bibnamefont {Lovorn}}, \bibinfo
  {author} {\bibfnamefont {E.}~\bibnamefont {Tutuc}}, \ and\ \bibinfo {author}
  {\bibfnamefont {A.}~\bibnamefont {MacDonald}},\ }\href {\doibase
  10.1103/PhysRevLett.121.026402} {\bibfield  {journal} {\bibinfo  {journal}
  {Physical Review Letters}\ }\textbf {\bibinfo {volume} {121}},\ \bibinfo
  {pages} {026402} (\bibinfo {year} {2018})}\BibitemShut {NoStop}%
\bibitem [{\citenamefont {Naik}\ and\ \citenamefont
  {Jain}(2018)}]{naikUltraflatbandsShearSolitons2018}%
  \BibitemOpen
  \bibfield  {author} {\bibinfo {author} {\bibfnamefont {M.~H.}\ \bibnamefont
  {Naik}}\ and\ \bibinfo {author} {\bibfnamefont {M.}~\bibnamefont {Jain}},\
  }\href {\doibase 10.1103/PhysRevLett.121.266401} {\bibfield  {journal}
  {\bibinfo  {journal} {Physical Review Letters}\ }\textbf {\bibinfo {volume}
  {121}},\ \bibinfo {pages} {266401} (\bibinfo {year} {2018})}\BibitemShut
  {NoStop}%
\bibitem [{\citenamefont {Regan}\ \emph {et~al.}(2020)\citenamefont {Regan},
  \citenamefont {Wang}, \citenamefont {Jin}, \citenamefont {Bakti~Utama},
  \citenamefont {Gao}, \citenamefont {Wei}, \citenamefont {Zhao}, \citenamefont
  {Zhao}, \citenamefont {Zhang}, \citenamefont {Yumigeta}, \citenamefont
  {Blei}, \citenamefont {Carlström}, \citenamefont {Watanabe}, \citenamefont
  {Taniguchi}, \citenamefont {Tongay}, \citenamefont {Crommie}, \citenamefont
  {Zettl},\ and\ \citenamefont {Wang}}]{reganMottGeneralizedWigner2020}%
  \BibitemOpen
  \bibfield  {author} {\bibinfo {author} {\bibfnamefont {E.~C.}\ \bibnamefont
  {Regan}}, \bibinfo {author} {\bibfnamefont {D.}~\bibnamefont {Wang}},
  \bibinfo {author} {\bibfnamefont {C.}~\bibnamefont {Jin}}, \bibinfo {author}
  {\bibfnamefont {M.~I.}\ \bibnamefont {Bakti~Utama}}, \bibinfo {author}
  {\bibfnamefont {B.}~\bibnamefont {Gao}}, \bibinfo {author} {\bibfnamefont
  {X.}~\bibnamefont {Wei}}, \bibinfo {author} {\bibfnamefont {S.}~\bibnamefont
  {Zhao}}, \bibinfo {author} {\bibfnamefont {W.}~\bibnamefont {Zhao}}, \bibinfo
  {author} {\bibfnamefont {Z.}~\bibnamefont {Zhang}}, \bibinfo {author}
  {\bibfnamefont {K.}~\bibnamefont {Yumigeta}}, \bibinfo {author}
  {\bibfnamefont {M.}~\bibnamefont {Blei}}, \bibinfo {author} {\bibfnamefont
  {J.~D.}\ \bibnamefont {Carlström}}, \bibinfo {author} {\bibfnamefont
  {K.}~\bibnamefont {Watanabe}}, \bibinfo {author} {\bibfnamefont
  {T.}~\bibnamefont {Taniguchi}}, \bibinfo {author} {\bibfnamefont
  {S.}~\bibnamefont {Tongay}}, \bibinfo {author} {\bibfnamefont
  {M.}~\bibnamefont {Crommie}}, \bibinfo {author} {\bibfnamefont
  {A.}~\bibnamefont {Zettl}}, \ and\ \bibinfo {author} {\bibfnamefont
  {F.}~\bibnamefont {Wang}},\ }\href {\doibase 10.1038/s41586-020-2092-4}
  {\bibfield  {journal} {\bibinfo  {journal} {Nature}\ }\textbf {\bibinfo
  {volume} {579}},\ \bibinfo {pages} {359} (\bibinfo {year}
  {2020})}\BibitemShut {NoStop}%
\bibitem [{\citenamefont {Li}\ \emph {et~al.}(2021)\citenamefont {Li},
  \citenamefont {Li}, \citenamefont {Regan}, \citenamefont {Wang},
  \citenamefont {Zhao}, \citenamefont {Kahn}, \citenamefont {Yumigeta},
  \citenamefont {Blei}, \citenamefont {Taniguchi}, \citenamefont {Watanabe},
  \citenamefont {Tongay}, \citenamefont {Zettl}, \citenamefont {Crommie},\ and\
  \citenamefont {Wang}}]{liImagingTwodimensionalGeneralized2021}%
  \BibitemOpen
  \bibfield  {author} {\bibinfo {author} {\bibfnamefont {H.}~\bibnamefont
  {Li}}, \bibinfo {author} {\bibfnamefont {S.}~\bibnamefont {Li}}, \bibinfo
  {author} {\bibfnamefont {E.~C.}\ \bibnamefont {Regan}}, \bibinfo {author}
  {\bibfnamefont {D.}~\bibnamefont {Wang}}, \bibinfo {author} {\bibfnamefont
  {W.}~\bibnamefont {Zhao}}, \bibinfo {author} {\bibfnamefont {S.}~\bibnamefont
  {Kahn}}, \bibinfo {author} {\bibfnamefont {K.}~\bibnamefont {Yumigeta}},
  \bibinfo {author} {\bibfnamefont {M.}~\bibnamefont {Blei}}, \bibinfo {author}
  {\bibfnamefont {T.}~\bibnamefont {Taniguchi}}, \bibinfo {author}
  {\bibfnamefont {K.}~\bibnamefont {Watanabe}}, \bibinfo {author}
  {\bibfnamefont {S.}~\bibnamefont {Tongay}}, \bibinfo {author} {\bibfnamefont
  {A.}~\bibnamefont {Zettl}}, \bibinfo {author} {\bibfnamefont {M.~F.}\
  \bibnamefont {Crommie}}, \ and\ \bibinfo {author} {\bibfnamefont
  {F.}~\bibnamefont {Wang}},\ }\href {\doibase 10.1038/s41586-021-03874-9}
  {\bibfield  {journal} {\bibinfo  {journal} {Nature}\ }\textbf {\bibinfo
  {volume} {597}},\ \bibinfo {pages} {650} (\bibinfo {year}
  {2021})}\BibitemShut {NoStop}%
\bibitem [{\citenamefont {Li}\ \emph {et~al.}(2024{\natexlab{a}})\citenamefont
  {Li}, \citenamefont {Xiang}, \citenamefont {Regan}, \citenamefont {Zhao},
  \citenamefont {Sailus}, \citenamefont {Banerjee}, \citenamefont {Taniguchi},
  \citenamefont {Watanabe}, \citenamefont {Tongay}, \citenamefont {Zettl},
  \citenamefont {Crommie},\ and\ \citenamefont
  {Wang}}]{liMappingChargeExcitations2024}%
  \BibitemOpen
  \bibfield  {author} {\bibinfo {author} {\bibfnamefont {H.}~\bibnamefont
  {Li}}, \bibinfo {author} {\bibfnamefont {Z.}~\bibnamefont {Xiang}}, \bibinfo
  {author} {\bibfnamefont {E.}~\bibnamefont {Regan}}, \bibinfo {author}
  {\bibfnamefont {W.}~\bibnamefont {Zhao}}, \bibinfo {author} {\bibfnamefont
  {R.}~\bibnamefont {Sailus}}, \bibinfo {author} {\bibfnamefont
  {R.}~\bibnamefont {Banerjee}}, \bibinfo {author} {\bibfnamefont
  {T.}~\bibnamefont {Taniguchi}}, \bibinfo {author} {\bibfnamefont
  {K.}~\bibnamefont {Watanabe}}, \bibinfo {author} {\bibfnamefont
  {S.}~\bibnamefont {Tongay}}, \bibinfo {author} {\bibfnamefont
  {A.}~\bibnamefont {Zettl}}, \bibinfo {author} {\bibfnamefont {M.~F.}\
  \bibnamefont {Crommie}}, \ and\ \bibinfo {author} {\bibfnamefont
  {F.}~\bibnamefont {Wang}},\ }\href {\doibase 10.1038/s41565-023-01594-x}
  {\bibfield  {journal} {\bibinfo  {journal} {Nature Nanotechnology}\ }\textbf
  {\bibinfo {volume} {19}},\ \bibinfo {pages} {618} (\bibinfo {year}
  {2024}{\natexlab{a}})}\BibitemShut {NoStop}%
\bibitem [{\citenamefont {Li}\ \emph {et~al.}(2024{\natexlab{b}})\citenamefont
  {Li}, \citenamefont {Xiang}, \citenamefont {Reddy}, \citenamefont {Devakul},
  \citenamefont {Sailus}, \citenamefont {Banerjee}, \citenamefont {Taniguchi},
  \citenamefont {Watanabe}, \citenamefont {Tongay}, \citenamefont {Zettl},
  \citenamefont {Fu}, \citenamefont {Crommie},\ and\ \citenamefont
  {Wang}}]{liWignerMolecularCrystals2024}%
  \BibitemOpen
  \bibfield  {author} {\bibinfo {author} {\bibfnamefont {H.}~\bibnamefont
  {Li}}, \bibinfo {author} {\bibfnamefont {Z.}~\bibnamefont {Xiang}}, \bibinfo
  {author} {\bibfnamefont {A.~P.}\ \bibnamefont {Reddy}}, \bibinfo {author}
  {\bibfnamefont {T.}~\bibnamefont {Devakul}}, \bibinfo {author} {\bibfnamefont
  {R.}~\bibnamefont {Sailus}}, \bibinfo {author} {\bibfnamefont
  {R.}~\bibnamefont {Banerjee}}, \bibinfo {author} {\bibfnamefont
  {T.}~\bibnamefont {Taniguchi}}, \bibinfo {author} {\bibfnamefont
  {K.}~\bibnamefont {Watanabe}}, \bibinfo {author} {\bibfnamefont
  {S.}~\bibnamefont {Tongay}}, \bibinfo {author} {\bibfnamefont
  {A.}~\bibnamefont {Zettl}}, \bibinfo {author} {\bibfnamefont
  {L.}~\bibnamefont {Fu}}, \bibinfo {author} {\bibfnamefont {M.~F.}\
  \bibnamefont {Crommie}}, \ and\ \bibinfo {author} {\bibfnamefont
  {F.}~\bibnamefont {Wang}},\ }\href {\doibase 10.1126/science.adk1348}
  {\bibfield  {journal} {\bibinfo  {journal} {Science}\ }\textbf {\bibinfo
  {volume} {385}},\ \bibinfo {pages} {86} (\bibinfo {year}
  {2024}{\natexlab{b}})}\BibitemShut {NoStop}%
\bibitem [{\citenamefont {Cai}\ \emph {et~al.}(2023{\natexlab{b}})\citenamefont
  {Cai}, \citenamefont {Anderson}, \citenamefont {Wang}, \citenamefont {Zhang},
  \citenamefont {Liu}, \citenamefont {Holtzmann}, \citenamefont {Zhang},
  \citenamefont {Fan}, \citenamefont {Taniguchi}, \citenamefont {Watanabe},
  \citenamefont {Ran}, \citenamefont {Cao}, \citenamefont {Fu}, \citenamefont
  {Xiao}, \citenamefont {Yao},\ and\ \citenamefont
  {Xu}}]{caiSignaturesFractionalQuantum2023a}%
  \BibitemOpen
  \bibfield  {author} {\bibinfo {author} {\bibfnamefont {J.}~\bibnamefont
  {Cai}}, \bibinfo {author} {\bibfnamefont {E.}~\bibnamefont {Anderson}},
  \bibinfo {author} {\bibfnamefont {C.}~\bibnamefont {Wang}}, \bibinfo {author}
  {\bibfnamefont {X.}~\bibnamefont {Zhang}}, \bibinfo {author} {\bibfnamefont
  {X.}~\bibnamefont {Liu}}, \bibinfo {author} {\bibfnamefont {W.}~\bibnamefont
  {Holtzmann}}, \bibinfo {author} {\bibfnamefont {Y.}~\bibnamefont {Zhang}},
  \bibinfo {author} {\bibfnamefont {F.}~\bibnamefont {Fan}}, \bibinfo {author}
  {\bibfnamefont {T.}~\bibnamefont {Taniguchi}}, \bibinfo {author}
  {\bibfnamefont {K.}~\bibnamefont {Watanabe}}, \bibinfo {author}
  {\bibfnamefont {Y.}~\bibnamefont {Ran}}, \bibinfo {author} {\bibfnamefont
  {T.}~\bibnamefont {Cao}}, \bibinfo {author} {\bibfnamefont {L.}~\bibnamefont
  {Fu}}, \bibinfo {author} {\bibfnamefont {D.}~\bibnamefont {Xiao}}, \bibinfo
  {author} {\bibfnamefont {W.}~\bibnamefont {Yao}}, \ and\ \bibinfo {author}
  {\bibfnamefont {X.}~\bibnamefont {Xu}},\ }\href {\doibase
  10.1038/s41586-023-06289-w} {\bibfield  {journal} {\bibinfo  {journal}
  {Nature}\ ,\ \bibinfo {pages} {1}} (\bibinfo {year}
  {2023}{\natexlab{b}})}\BibitemShut {NoStop}%
\bibitem [{\citenamefont {Zeng}\ \emph
  {et~al.}(2023{\natexlab{b}})\citenamefont {Zeng}, \citenamefont {Xia},
  \citenamefont {Kang}, \citenamefont {Zhu}, \citenamefont {Knüppel},
  \citenamefont {Vaswani}, \citenamefont {Watanabe}, \citenamefont {Taniguchi},
  \citenamefont {Mak},\ and\ \citenamefont
  {Shan}}]{zengThermodynamicEvidenceFractional2023a}%
  \BibitemOpen
  \bibfield  {author} {\bibinfo {author} {\bibfnamefont {Y.}~\bibnamefont
  {Zeng}}, \bibinfo {author} {\bibfnamefont {Z.}~\bibnamefont {Xia}}, \bibinfo
  {author} {\bibfnamefont {K.}~\bibnamefont {Kang}}, \bibinfo {author}
  {\bibfnamefont {J.}~\bibnamefont {Zhu}}, \bibinfo {author} {\bibfnamefont
  {P.}~\bibnamefont {Knüppel}}, \bibinfo {author} {\bibfnamefont
  {C.}~\bibnamefont {Vaswani}}, \bibinfo {author} {\bibfnamefont
  {K.}~\bibnamefont {Watanabe}}, \bibinfo {author} {\bibfnamefont
  {T.}~\bibnamefont {Taniguchi}}, \bibinfo {author} {\bibfnamefont {K.~F.}\
  \bibnamefont {Mak}}, \ and\ \bibinfo {author} {\bibfnamefont
  {J.}~\bibnamefont {Shan}},\ }\href {\doibase 10.1038/s41586-023-06452-3}
  {\bibfield  {journal} {\bibinfo  {journal} {Nature}\ ,\ \bibinfo {pages} {1}}
  (\bibinfo {year} {2023}{\natexlab{b}})}\BibitemShut {NoStop}%
\bibitem [{\citenamefont {Xu}\ \emph {et~al.}(2023)\citenamefont {Xu},
  \citenamefont {Sun}, \citenamefont {Jia}, \citenamefont {Liu}, \citenamefont
  {Xu}, \citenamefont {Li}, \citenamefont {Gu}, \citenamefont {Watanabe},
  \citenamefont {Taniguchi}, \citenamefont {Tong}, \citenamefont {Jia},
  \citenamefont {Shi}, \citenamefont {Jiang}, \citenamefont {Zhang},
  \citenamefont {Liu},\ and\ \citenamefont
  {Li}}]{xuObservationIntegerFractional2023}%
  \BibitemOpen
  \bibfield  {author} {\bibinfo {author} {\bibfnamefont {F.}~\bibnamefont
  {Xu}}, \bibinfo {author} {\bibfnamefont {Z.}~\bibnamefont {Sun}}, \bibinfo
  {author} {\bibfnamefont {T.}~\bibnamefont {Jia}}, \bibinfo {author}
  {\bibfnamefont {C.}~\bibnamefont {Liu}}, \bibinfo {author} {\bibfnamefont
  {C.}~\bibnamefont {Xu}}, \bibinfo {author} {\bibfnamefont {C.}~\bibnamefont
  {Li}}, \bibinfo {author} {\bibfnamefont {Y.}~\bibnamefont {Gu}}, \bibinfo
  {author} {\bibfnamefont {K.}~\bibnamefont {Watanabe}}, \bibinfo {author}
  {\bibfnamefont {T.}~\bibnamefont {Taniguchi}}, \bibinfo {author}
  {\bibfnamefont {B.}~\bibnamefont {Tong}}, \bibinfo {author} {\bibfnamefont
  {J.}~\bibnamefont {Jia}}, \bibinfo {author} {\bibfnamefont {Z.}~\bibnamefont
  {Shi}}, \bibinfo {author} {\bibfnamefont {S.}~\bibnamefont {Jiang}}, \bibinfo
  {author} {\bibfnamefont {Y.}~\bibnamefont {Zhang}}, \bibinfo {author}
  {\bibfnamefont {X.}~\bibnamefont {Liu}}, \ and\ \bibinfo {author}
  {\bibfnamefont {T.}~\bibnamefont {Li}},\ }\href {\doibase
  10.1103/PhysRevX.13.031037} {\bibfield  {journal} {\bibinfo  {journal}
  {Physical Review X}\ }\textbf {\bibinfo {volume} {13}},\ \bibinfo {pages}
  {031037} (\bibinfo {year} {2023})}\BibitemShut {NoStop}%
\bibitem [{\citenamefont {Foutty}\ \emph
  {et~al.}(2024{\natexlab{a}})\citenamefont {Foutty}, \citenamefont {Kometter},
  \citenamefont {Devakul}, \citenamefont {Reddy}, \citenamefont {Watanabe},
  \citenamefont {Taniguchi}, \citenamefont {Fu},\ and\ \citenamefont
  {Feldman}}]{fouttyMappingTwisttunedMultiband2024}%
  \BibitemOpen
  \bibfield  {author} {\bibinfo {author} {\bibfnamefont {B.~A.}\ \bibnamefont
  {Foutty}}, \bibinfo {author} {\bibfnamefont {C.~R.}\ \bibnamefont
  {Kometter}}, \bibinfo {author} {\bibfnamefont {T.}~\bibnamefont {Devakul}},
  \bibinfo {author} {\bibfnamefont {A.~P.}\ \bibnamefont {Reddy}}, \bibinfo
  {author} {\bibfnamefont {K.}~\bibnamefont {Watanabe}}, \bibinfo {author}
  {\bibfnamefont {T.}~\bibnamefont {Taniguchi}}, \bibinfo {author}
  {\bibfnamefont {L.}~\bibnamefont {Fu}}, \ and\ \bibinfo {author}
  {\bibfnamefont {B.~E.}\ \bibnamefont {Feldman}},\ }\href {\doibase
  10.1126/science.adi4728} {\bibfield  {journal} {\bibinfo  {journal}
  {Science}\ }\textbf {\bibinfo {volume} {384}},\ \bibinfo {pages} {343}
  (\bibinfo {year} {2024}{\natexlab{a}})}\BibitemShut {NoStop}%
\bibitem [{\citenamefont {Kang}\ \emph {et~al.}(2024)\citenamefont {Kang},
  \citenamefont {Shen}, \citenamefont {Qiu}, \citenamefont {Zeng},
  \citenamefont {Xia}, \citenamefont {Watanabe}, \citenamefont {Taniguchi},
  \citenamefont {Shan},\ and\ \citenamefont
  {Mak}}]{kangEvidenceFractionalQuantum2024}%
  \BibitemOpen
  \bibfield  {author} {\bibinfo {author} {\bibfnamefont {K.}~\bibnamefont
  {Kang}}, \bibinfo {author} {\bibfnamefont {B.}~\bibnamefont {Shen}}, \bibinfo
  {author} {\bibfnamefont {Y.}~\bibnamefont {Qiu}}, \bibinfo {author}
  {\bibfnamefont {Y.}~\bibnamefont {Zeng}}, \bibinfo {author} {\bibfnamefont
  {Z.}~\bibnamefont {Xia}}, \bibinfo {author} {\bibfnamefont {K.}~\bibnamefont
  {Watanabe}}, \bibinfo {author} {\bibfnamefont {T.}~\bibnamefont {Taniguchi}},
  \bibinfo {author} {\bibfnamefont {J.}~\bibnamefont {Shan}}, \ and\ \bibinfo
  {author} {\bibfnamefont {K.~F.}\ \bibnamefont {Mak}},\ }\href {\doibase
  10.1038/s41586-024-07214-5} {\bibfield  {journal} {\bibinfo  {journal}
  {Nature}\ }\textbf {\bibinfo {volume} {628}},\ \bibinfo {pages} {522}
  (\bibinfo {year} {2024})}\BibitemShut {NoStop}%
\bibitem [{\citenamefont {Xia}\ \emph {et~al.}(2025)\citenamefont {Xia},
  \citenamefont {Han}, \citenamefont {Watanabe}, \citenamefont {Taniguchi},
  \citenamefont {Shan},\ and\ \citenamefont
  {Mak}}]{xiaSuperconductivityTwistedBilayer2025}%
  \BibitemOpen
  \bibfield  {author} {\bibinfo {author} {\bibfnamefont {Y.}~\bibnamefont
  {Xia}}, \bibinfo {author} {\bibfnamefont {Z.}~\bibnamefont {Han}}, \bibinfo
  {author} {\bibfnamefont {K.}~\bibnamefont {Watanabe}}, \bibinfo {author}
  {\bibfnamefont {T.}~\bibnamefont {Taniguchi}}, \bibinfo {author}
  {\bibfnamefont {J.}~\bibnamefont {Shan}}, \ and\ \bibinfo {author}
  {\bibfnamefont {K.~F.}\ \bibnamefont {Mak}},\ }\href {\doibase
  10.1038/s41586-024-08116-2} {\bibfield  {journal} {\bibinfo  {journal}
  {Nature}\ }\textbf {\bibinfo {volume} {637}},\ \bibinfo {pages} {833}
  (\bibinfo {year} {2025})}\BibitemShut {NoStop}%
\bibitem [{\citenamefont {Guo}\ \emph {et~al.}(2025)\citenamefont {Guo},
  \citenamefont {Pack}, \citenamefont {Swann}, \citenamefont {Holtzman},
  \citenamefont {Cothrine}, \citenamefont {Watanabe}, \citenamefont
  {Taniguchi}, \citenamefont {Mandrus}, \citenamefont {Barmak}, \citenamefont
  {Hone}, \citenamefont {Millis}, \citenamefont {Pasupathy},\ and\
  \citenamefont {Dean}}]{guoSuperconductivity50degTwisted2025}%
  \BibitemOpen
  \bibfield  {author} {\bibinfo {author} {\bibfnamefont {Y.}~\bibnamefont
  {Guo}}, \bibinfo {author} {\bibfnamefont {J.}~\bibnamefont {Pack}}, \bibinfo
  {author} {\bibfnamefont {J.}~\bibnamefont {Swann}}, \bibinfo {author}
  {\bibfnamefont {L.}~\bibnamefont {Holtzman}}, \bibinfo {author}
  {\bibfnamefont {M.}~\bibnamefont {Cothrine}}, \bibinfo {author}
  {\bibfnamefont {K.}~\bibnamefont {Watanabe}}, \bibinfo {author}
  {\bibfnamefont {T.}~\bibnamefont {Taniguchi}}, \bibinfo {author}
  {\bibfnamefont {D.~G.}\ \bibnamefont {Mandrus}}, \bibinfo {author}
  {\bibfnamefont {K.}~\bibnamefont {Barmak}}, \bibinfo {author} {\bibfnamefont
  {J.}~\bibnamefont {Hone}}, \bibinfo {author} {\bibfnamefont {A.~J.}\
  \bibnamefont {Millis}}, \bibinfo {author} {\bibfnamefont {A.}~\bibnamefont
  {Pasupathy}}, \ and\ \bibinfo {author} {\bibfnamefont {C.~R.}\ \bibnamefont
  {Dean}},\ }\href {\doibase 10.1038/s41586-024-08381-1} {\bibfield  {journal}
  {\bibinfo  {journal} {Nature}\ }\textbf {\bibinfo {volume} {637}},\ \bibinfo
  {pages} {839} (\bibinfo {year} {2025})}\BibitemShut {NoStop}%
\bibitem [{\citenamefont {Wu}\ \emph {et~al.}(2023)\citenamefont {Wu},
  \citenamefont {Xu}, \citenamefont {Wang}, \citenamefont {Chu}, \citenamefont
  {Li}, \citenamefont {Tang}, \citenamefont {Liu}, \citenamefont {Tian},
  \citenamefont {Ji}, \citenamefont {Liu}, \citenamefont {Yuan}, \citenamefont
  {Huang}, \citenamefont {Zhao}, \citenamefont {Zan}, \citenamefont {Watanabe},
  \citenamefont {Taniguchi}, \citenamefont {Shi}, \citenamefont {Gu},
  \citenamefont {Xu}, \citenamefont {Xian}, \citenamefont {Yang}, \citenamefont
  {Du},\ and\ \citenamefont {Zhang}}]{wuGiantCorrelatedGap2023}%
  \BibitemOpen
  \bibfield  {author} {\bibinfo {author} {\bibfnamefont {F.}~\bibnamefont
  {Wu}}, \bibinfo {author} {\bibfnamefont {Q.}~\bibnamefont {Xu}}, \bibinfo
  {author} {\bibfnamefont {Q.}~\bibnamefont {Wang}}, \bibinfo {author}
  {\bibfnamefont {Y.}~\bibnamefont {Chu}}, \bibinfo {author} {\bibfnamefont
  {L.}~\bibnamefont {Li}}, \bibinfo {author} {\bibfnamefont {J.}~\bibnamefont
  {Tang}}, \bibinfo {author} {\bibfnamefont {J.}~\bibnamefont {Liu}}, \bibinfo
  {author} {\bibfnamefont {J.}~\bibnamefont {Tian}}, \bibinfo {author}
  {\bibfnamefont {Y.}~\bibnamefont {Ji}}, \bibinfo {author} {\bibfnamefont
  {L.}~\bibnamefont {Liu}}, \bibinfo {author} {\bibfnamefont {Y.}~\bibnamefont
  {Yuan}}, \bibinfo {author} {\bibfnamefont {Z.}~\bibnamefont {Huang}},
  \bibinfo {author} {\bibfnamefont {J.}~\bibnamefont {Zhao}}, \bibinfo {author}
  {\bibfnamefont {X.}~\bibnamefont {Zan}}, \bibinfo {author} {\bibfnamefont
  {K.}~\bibnamefont {Watanabe}}, \bibinfo {author} {\bibfnamefont
  {T.}~\bibnamefont {Taniguchi}}, \bibinfo {author} {\bibfnamefont
  {D.}~\bibnamefont {Shi}}, \bibinfo {author} {\bibfnamefont {G.}~\bibnamefont
  {Gu}}, \bibinfo {author} {\bibfnamefont {Y.}~\bibnamefont {Xu}}, \bibinfo
  {author} {\bibfnamefont {L.}~\bibnamefont {Xian}}, \bibinfo {author}
  {\bibfnamefont {W.}~\bibnamefont {Yang}}, \bibinfo {author} {\bibfnamefont
  {L.}~\bibnamefont {Du}}, \ and\ \bibinfo {author} {\bibfnamefont
  {G.}~\bibnamefont {Zhang}},\ }\href {\doibase 10.1103/PhysRevLett.131.256201}
  {\bibfield  {journal} {\bibinfo  {journal} {Physical Review Letters}\
  }\textbf {\bibinfo {volume} {131}},\ \bibinfo {pages} {256201} (\bibinfo
  {year} {2023})}\BibitemShut {NoStop}%
\bibitem [{Sup()}]{SupplementalMaterial}%
  \BibitemOpen
  \href@noop {} {}\bibinfo {note} {See Supplemental Material at [URL will be
  inserted by publisher] for continuum-model calculations, device fabrication
  and measurement controls, electrostatic and filling calibration, and analysis
  of the conductance features.}\BibitemShut {Stop}%
\bibitem [{\citenamefont {Wang}\ \emph {et~al.}(2019)\citenamefont {Wang},
  \citenamefont {Kim}, \citenamefont {Wu}, \citenamefont {Martinez},
  \citenamefont {Song}, \citenamefont {Yang}, \citenamefont {Zhao},
  \citenamefont {Mkhoyan}, \citenamefont {Jeong},\ and\ \citenamefont
  {Chhowalla}}]{wangVanWaalsContacts2019}%
  \BibitemOpen
  \bibfield  {author} {\bibinfo {author} {\bibfnamefont {Y.}~\bibnamefont
  {Wang}}, \bibinfo {author} {\bibfnamefont {J.~C.}\ \bibnamefont {Kim}},
  \bibinfo {author} {\bibfnamefont {R.~J.}\ \bibnamefont {Wu}}, \bibinfo
  {author} {\bibfnamefont {J.}~\bibnamefont {Martinez}}, \bibinfo {author}
  {\bibfnamefont {X.}~\bibnamefont {Song}}, \bibinfo {author} {\bibfnamefont
  {J.}~\bibnamefont {Yang}}, \bibinfo {author} {\bibfnamefont {F.}~\bibnamefont
  {Zhao}}, \bibinfo {author} {\bibfnamefont {A.}~\bibnamefont {Mkhoyan}},
  \bibinfo {author} {\bibfnamefont {H.~Y.}\ \bibnamefont {Jeong}}, \ and\
  \bibinfo {author} {\bibfnamefont {M.}~\bibnamefont {Chhowalla}},\ }\href
  {\doibase 10.1038/s41586-019-1052-3} {\bibfield  {journal} {\bibinfo
  {journal} {Nature}\ }\textbf {\bibinfo {volume} {568}},\ \bibinfo {pages}
  {70} (\bibinfo {year} {2019})},\ \bibinfo {note} {number: 7750}\BibitemShut
  {NoStop}%
\bibitem [{\citenamefont {Wang}\ \emph {et~al.}(2022)\citenamefont {Wang},
  \citenamefont {Kim}, \citenamefont {Li}, \citenamefont {Ma}, \citenamefont
  {Hong}, \citenamefont {Kim}, \citenamefont {Shin}, \citenamefont {Jeong},\
  and\ \citenamefont {Chhowalla}}]{wangPtypeElectricalContacts2022}%
  \BibitemOpen
  \bibfield  {author} {\bibinfo {author} {\bibfnamefont {Y.}~\bibnamefont
  {Wang}}, \bibinfo {author} {\bibfnamefont {J.~C.}\ \bibnamefont {Kim}},
  \bibinfo {author} {\bibfnamefont {Y.}~\bibnamefont {Li}}, \bibinfo {author}
  {\bibfnamefont {K.~Y.}\ \bibnamefont {Ma}}, \bibinfo {author} {\bibfnamefont
  {S.}~\bibnamefont {Hong}}, \bibinfo {author} {\bibfnamefont {M.}~\bibnamefont
  {Kim}}, \bibinfo {author} {\bibfnamefont {H.~S.}\ \bibnamefont {Shin}},
  \bibinfo {author} {\bibfnamefont {H.~Y.}\ \bibnamefont {Jeong}}, \ and\
  \bibinfo {author} {\bibfnamefont {M.}~\bibnamefont {Chhowalla}},\ }\href
  {\doibase 10.1038/s41586-022-05134-w} {\bibfield  {journal} {\bibinfo
  {journal} {Nature}\ }\textbf {\bibinfo {volume} {610}},\ \bibinfo {pages}
  {61} (\bibinfo {year} {2022})},\ \bibinfo {note} {number: 7930}\BibitemShut
  {NoStop}%
\bibitem [{\citenamefont {Wang}\ \emph {et~al.}(2024)\citenamefont {Wang},
  \citenamefont {Sarkar}, \citenamefont {Yan},\ and\ \citenamefont
  {Chhowalla}}]{wangCriticalChallengesDevelopment2024}%
  \BibitemOpen
  \bibfield  {author} {\bibinfo {author} {\bibfnamefont {Y.}~\bibnamefont
  {Wang}}, \bibinfo {author} {\bibfnamefont {S.}~\bibnamefont {Sarkar}},
  \bibinfo {author} {\bibfnamefont {H.}~\bibnamefont {Yan}}, \ and\ \bibinfo
  {author} {\bibfnamefont {M.}~\bibnamefont {Chhowalla}},\ }\href {\doibase
  10.1038/s41928-024-01210-3} {\bibfield  {journal} {\bibinfo  {journal}
  {Nature Electronics}\ ,\ \bibinfo {pages} {1}} (\bibinfo {year}
  {2024})}\BibitemShut {NoStop}%
\bibitem [{\citenamefont {Pack}\ \emph {et~al.}(2024)\citenamefont {Pack},
  \citenamefont {Guo}, \citenamefont {Liu}, \citenamefont {Jessen},
  \citenamefont {Holtzman}, \citenamefont {Liu}, \citenamefont {Cothrine},
  \citenamefont {Watanabe}, \citenamefont {Taniguchi}, \citenamefont {Mandrus},
  \citenamefont {Barmak}, \citenamefont {Hone},\ and\ \citenamefont
  {Dean}}]{packChargetransferContactsMeasurement2024}%
  \BibitemOpen
  \bibfield  {author} {\bibinfo {author} {\bibfnamefont {J.}~\bibnamefont
  {Pack}}, \bibinfo {author} {\bibfnamefont {Y.}~\bibnamefont {Guo}}, \bibinfo
  {author} {\bibfnamefont {Z.}~\bibnamefont {Liu}}, \bibinfo {author}
  {\bibfnamefont {B.~S.}\ \bibnamefont {Jessen}}, \bibinfo {author}
  {\bibfnamefont {L.}~\bibnamefont {Holtzman}}, \bibinfo {author}
  {\bibfnamefont {S.}~\bibnamefont {Liu}}, \bibinfo {author} {\bibfnamefont
  {M.}~\bibnamefont {Cothrine}}, \bibinfo {author} {\bibfnamefont
  {K.}~\bibnamefont {Watanabe}}, \bibinfo {author} {\bibfnamefont
  {T.}~\bibnamefont {Taniguchi}}, \bibinfo {author} {\bibfnamefont {D.~G.}\
  \bibnamefont {Mandrus}}, \bibinfo {author} {\bibfnamefont {K.}~\bibnamefont
  {Barmak}}, \bibinfo {author} {\bibfnamefont {J.}~\bibnamefont {Hone}}, \ and\
  \bibinfo {author} {\bibfnamefont {C.~R.}\ \bibnamefont {Dean}},\ }\href
  {\doibase 10.1038/s41565-024-01702-5} {\bibfield  {journal} {\bibinfo
  {journal} {Nature Nanotechnology}\ }\textbf {\bibinfo {volume} {19}},\
  \bibinfo {pages} {948} (\bibinfo {year} {2024})}\BibitemShut {NoStop}%
\bibitem [{\citenamefont {Uri}\ \emph {et~al.}(2023)\citenamefont {Uri},
  \citenamefont {de~la Barrera}, \citenamefont {Randeria}, \citenamefont
  {Rodan-Legrain}, \citenamefont {Devakul}, \citenamefont {Crowley},
  \citenamefont {Paul}, \citenamefont {Watanabe}, \citenamefont {Taniguchi},
  \citenamefont {Lifshitz}, \citenamefont {Fu}, \citenamefont {Ashoori},\ and\
  \citenamefont
  {Jarillo-Herrero}}]{uriSuperconductivityStrongInteractions2023}%
  \BibitemOpen
  \bibfield  {author} {\bibinfo {author} {\bibfnamefont {A.}~\bibnamefont
  {Uri}}, \bibinfo {author} {\bibfnamefont {S.~C.}\ \bibnamefont {de~la
  Barrera}}, \bibinfo {author} {\bibfnamefont {M.~T.}\ \bibnamefont
  {Randeria}}, \bibinfo {author} {\bibfnamefont {D.}~\bibnamefont
  {Rodan-Legrain}}, \bibinfo {author} {\bibfnamefont {T.}~\bibnamefont
  {Devakul}}, \bibinfo {author} {\bibfnamefont {P.~J.~D.}\ \bibnamefont
  {Crowley}}, \bibinfo {author} {\bibfnamefont {N.}~\bibnamefont {Paul}},
  \bibinfo {author} {\bibfnamefont {K.}~\bibnamefont {Watanabe}}, \bibinfo
  {author} {\bibfnamefont {T.}~\bibnamefont {Taniguchi}}, \bibinfo {author}
  {\bibfnamefont {R.}~\bibnamefont {Lifshitz}}, \bibinfo {author}
  {\bibfnamefont {L.}~\bibnamefont {Fu}}, \bibinfo {author} {\bibfnamefont
  {R.~C.}\ \bibnamefont {Ashoori}}, \ and\ \bibinfo {author} {\bibfnamefont
  {P.}~\bibnamefont {Jarillo-Herrero}},\ }\href {\doibase
  10.1038/s41586-023-06294-z} {\bibfield  {journal} {\bibinfo  {journal}
  {Nature}\ }\textbf {\bibinfo {volume} {620}},\ \bibinfo {pages} {762}
  (\bibinfo {year} {2023})}\BibitemShut {NoStop}%
\bibitem [{\citenamefont {Yoo}\ \emph {et~al.}(2019)\citenamefont {Yoo},
  \citenamefont {Engelke}, \citenamefont {Carr}, \citenamefont {Fang},
  \citenamefont {Zhang}, \citenamefont {Cazeaux}, \citenamefont {Sung},
  \citenamefont {Hovden}, \citenamefont {Tsen}, \citenamefont {Taniguchi},
  \citenamefont {Watanabe}, \citenamefont {Yi}, \citenamefont {Kim},
  \citenamefont {Luskin}, \citenamefont {Tadmor}, \citenamefont {Kaxiras},\
  and\ \citenamefont {Kim}}]{yooAtomicElectronicReconstruction2019a}%
  \BibitemOpen
  \bibfield  {author} {\bibinfo {author} {\bibfnamefont {H.}~\bibnamefont
  {Yoo}}, \bibinfo {author} {\bibfnamefont {R.}~\bibnamefont {Engelke}},
  \bibinfo {author} {\bibfnamefont {S.}~\bibnamefont {Carr}}, \bibinfo {author}
  {\bibfnamefont {S.}~\bibnamefont {Fang}}, \bibinfo {author} {\bibfnamefont
  {K.}~\bibnamefont {Zhang}}, \bibinfo {author} {\bibfnamefont
  {P.}~\bibnamefont {Cazeaux}}, \bibinfo {author} {\bibfnamefont {S.~H.}\
  \bibnamefont {Sung}}, \bibinfo {author} {\bibfnamefont {R.}~\bibnamefont
  {Hovden}}, \bibinfo {author} {\bibfnamefont {A.~W.}\ \bibnamefont {Tsen}},
  \bibinfo {author} {\bibfnamefont {T.}~\bibnamefont {Taniguchi}}, \bibinfo
  {author} {\bibfnamefont {K.}~\bibnamefont {Watanabe}}, \bibinfo {author}
  {\bibfnamefont {G.-C.}\ \bibnamefont {Yi}}, \bibinfo {author} {\bibfnamefont
  {M.}~\bibnamefont {Kim}}, \bibinfo {author} {\bibfnamefont {M.}~\bibnamefont
  {Luskin}}, \bibinfo {author} {\bibfnamefont {E.~B.}\ \bibnamefont {Tadmor}},
  \bibinfo {author} {\bibfnamefont {E.}~\bibnamefont {Kaxiras}}, \ and\
  \bibinfo {author} {\bibfnamefont {P.}~\bibnamefont {Kim}},\ }\href {\doibase
  10.1038/s41563-019-0346-z} {\bibfield  {journal} {\bibinfo  {journal} {Nature
  Materials}\ }\textbf {\bibinfo {volume} {18}},\ \bibinfo {pages} {448}
  (\bibinfo {year} {2019})}\BibitemShut {NoStop}%
\bibitem [{\citenamefont {Wong}\ \emph
  {et~al.}(2020{\natexlab{a}})\citenamefont {Wong}, \citenamefont {Nuckolls},
  \citenamefont {Oh}, \citenamefont {Lian}, \citenamefont {Xie}, \citenamefont
  {Jeon}, \citenamefont {Watanabe}, \citenamefont {Taniguchi}, \citenamefont
  {Bernevig},\ and\ \citenamefont
  {Yazdani}}]{wongCascadeElectronicTransitions2020a}%
  \BibitemOpen
  \bibfield  {author} {\bibinfo {author} {\bibfnamefont {D.}~\bibnamefont
  {Wong}}, \bibinfo {author} {\bibfnamefont {K.~P.}\ \bibnamefont {Nuckolls}},
  \bibinfo {author} {\bibfnamefont {M.}~\bibnamefont {Oh}}, \bibinfo {author}
  {\bibfnamefont {B.}~\bibnamefont {Lian}}, \bibinfo {author} {\bibfnamefont
  {Y.}~\bibnamefont {Xie}}, \bibinfo {author} {\bibfnamefont {S.}~\bibnamefont
  {Jeon}}, \bibinfo {author} {\bibfnamefont {K.}~\bibnamefont {Watanabe}},
  \bibinfo {author} {\bibfnamefont {T.}~\bibnamefont {Taniguchi}}, \bibinfo
  {author} {\bibfnamefont {B.~A.}\ \bibnamefont {Bernevig}}, \ and\ \bibinfo
  {author} {\bibfnamefont {A.}~\bibnamefont {Yazdani}},\ }\href {\doibase
  10.1038/s41586-020-2339-0} {\bibfield  {journal} {\bibinfo  {journal}
  {Nature}\ }\textbf {\bibinfo {volume} {582}},\ \bibinfo {pages} {198}
  (\bibinfo {year} {2020}{\natexlab{a}})}\BibitemShut {NoStop}%
\bibitem [{\citenamefont {Kazmierczak}\ \emph {et~al.}(2021)\citenamefont
  {Kazmierczak}, \citenamefont {Van~Winkle}, \citenamefont {Ophus},
  \citenamefont {Bustillo}, \citenamefont {Carr}, \citenamefont {Brown},
  \citenamefont {Ciston}, \citenamefont {Taniguchi}, \citenamefont {Watanabe},\
  and\ \citenamefont {Bediako}}]{kazmierczakStrainFieldsTwisted2021}%
  \BibitemOpen
  \bibfield  {author} {\bibinfo {author} {\bibfnamefont {N.~P.}\ \bibnamefont
  {Kazmierczak}}, \bibinfo {author} {\bibfnamefont {M.}~\bibnamefont
  {Van~Winkle}}, \bibinfo {author} {\bibfnamefont {C.}~\bibnamefont {Ophus}},
  \bibinfo {author} {\bibfnamefont {K.~C.}\ \bibnamefont {Bustillo}}, \bibinfo
  {author} {\bibfnamefont {S.}~\bibnamefont {Carr}}, \bibinfo {author}
  {\bibfnamefont {H.~G.}\ \bibnamefont {Brown}}, \bibinfo {author}
  {\bibfnamefont {J.}~\bibnamefont {Ciston}}, \bibinfo {author} {\bibfnamefont
  {T.}~\bibnamefont {Taniguchi}}, \bibinfo {author} {\bibfnamefont
  {K.}~\bibnamefont {Watanabe}}, \ and\ \bibinfo {author} {\bibfnamefont
  {D.~K.}\ \bibnamefont {Bediako}},\ }\href {\doibase
  10.1038/s41563-021-00973-w} {\bibfield  {journal} {\bibinfo  {journal}
  {Nature Materials}\ }\textbf {\bibinfo {volume} {20}},\ \bibinfo {pages}
  {956} (\bibinfo {year} {2021})}\BibitemShut {NoStop}%
\bibitem [{\citenamefont {Wong}\ \emph
  {et~al.}(2020{\natexlab{b}})\citenamefont {Wong}, \citenamefont {Nuckolls},
  \citenamefont {Oh}, \citenamefont {Lian}, \citenamefont {Xie}, \citenamefont
  {Jeon}, \citenamefont {Watanabe}, \citenamefont {Taniguchi}, \citenamefont
  {Bernevig},\ and\ \citenamefont
  {Yazdani}}]{wongCascadeElectronicTransitions2020}%
  \BibitemOpen
  \bibfield  {author} {\bibinfo {author} {\bibfnamefont {D.}~\bibnamefont
  {Wong}}, \bibinfo {author} {\bibfnamefont {K.~P.}\ \bibnamefont {Nuckolls}},
  \bibinfo {author} {\bibfnamefont {M.}~\bibnamefont {Oh}}, \bibinfo {author}
  {\bibfnamefont {B.}~\bibnamefont {Lian}}, \bibinfo {author} {\bibfnamefont
  {Y.}~\bibnamefont {Xie}}, \bibinfo {author} {\bibfnamefont {S.}~\bibnamefont
  {Jeon}}, \bibinfo {author} {\bibfnamefont {K.}~\bibnamefont {Watanabe}},
  \bibinfo {author} {\bibfnamefont {T.}~\bibnamefont {Taniguchi}}, \bibinfo
  {author} {\bibfnamefont {B.~A.}\ \bibnamefont {Bernevig}}, \ and\ \bibinfo
  {author} {\bibfnamefont {A.}~\bibnamefont {Yazdani}},\ }\href {\doibase
  10.1038/s41586-020-2339-0} {\bibfield  {journal} {\bibinfo  {journal}
  {Nature}\ }\textbf {\bibinfo {volume} {582}},\ \bibinfo {pages} {198}
  (\bibinfo {year} {2020}{\natexlab{b}})}\BibitemShut {NoStop}%
\bibitem [{\citenamefont {Xie}\ \emph {et~al.}(2021)\citenamefont {Xie},
  \citenamefont {Pierce}, \citenamefont {Park}, \citenamefont {Parker},
  \citenamefont {Khalaf}, \citenamefont {Ledwith}, \citenamefont {Cao},
  \citenamefont {Lee}, \citenamefont {Chen}, \citenamefont {Forrester},
  \citenamefont {Watanabe}, \citenamefont {Taniguchi}, \citenamefont
  {Vishwanath}, \citenamefont {Jarillo-Herrero},\ and\ \citenamefont
  {Yacoby}}]{xieFractionalChernInsulators2021}%
  \BibitemOpen
  \bibfield  {author} {\bibinfo {author} {\bibfnamefont {Y.}~\bibnamefont
  {Xie}}, \bibinfo {author} {\bibfnamefont {A.~T.}\ \bibnamefont {Pierce}},
  \bibinfo {author} {\bibfnamefont {J.~M.}\ \bibnamefont {Park}}, \bibinfo
  {author} {\bibfnamefont {D.~E.}\ \bibnamefont {Parker}}, \bibinfo {author}
  {\bibfnamefont {E.}~\bibnamefont {Khalaf}}, \bibinfo {author} {\bibfnamefont
  {P.}~\bibnamefont {Ledwith}}, \bibinfo {author} {\bibfnamefont
  {Y.}~\bibnamefont {Cao}}, \bibinfo {author} {\bibfnamefont {S.~H.}\
  \bibnamefont {Lee}}, \bibinfo {author} {\bibfnamefont {S.}~\bibnamefont
  {Chen}}, \bibinfo {author} {\bibfnamefont {P.~R.}\ \bibnamefont {Forrester}},
  \bibinfo {author} {\bibfnamefont {K.}~\bibnamefont {Watanabe}}, \bibinfo
  {author} {\bibfnamefont {T.}~\bibnamefont {Taniguchi}}, \bibinfo {author}
  {\bibfnamefont {A.}~\bibnamefont {Vishwanath}}, \bibinfo {author}
  {\bibfnamefont {P.}~\bibnamefont {Jarillo-Herrero}}, \ and\ \bibinfo {author}
  {\bibfnamefont {A.}~\bibnamefont {Yacoby}},\ }\href {\doibase
  10.1038/s41586-021-04002-3} {\bibfield  {journal} {\bibinfo  {journal}
  {Nature}\ }\textbf {\bibinfo {volume} {600}},\ \bibinfo {pages} {439}
  (\bibinfo {year} {2021})}\BibitemShut {NoStop}%
\bibitem [{\citenamefont {Foutty}\ \emph
  {et~al.}(2024{\natexlab{b}})\citenamefont {Foutty}, \citenamefont {Kometter},
  \citenamefont {Devakul}, \citenamefont {Reddy}, \citenamefont {Watanabe},
  \citenamefont {Taniguchi}, \citenamefont {Fu},\ and\ \citenamefont
  {Feldman}}]{fouttyDatasetMappingTwisttuned2024}%
  \BibitemOpen
  \bibfield  {author} {\bibinfo {author} {\bibfnamefont {B.~A.}\ \bibnamefont
  {Foutty}}, \bibinfo {author} {\bibfnamefont {C.~R.}\ \bibnamefont
  {Kometter}}, \bibinfo {author} {\bibfnamefont {T.}~\bibnamefont {Devakul}},
  \bibinfo {author} {\bibfnamefont {A.~P.}\ \bibnamefont {Reddy}}, \bibinfo
  {author} {\bibfnamefont {K.}~\bibnamefont {Watanabe}}, \bibinfo {author}
  {\bibfnamefont {T.}~\bibnamefont {Taniguchi}}, \bibinfo {author}
  {\bibfnamefont {L.}~\bibnamefont {Fu}}, \ and\ \bibinfo {author}
  {\bibfnamefont {B.~E.}\ \bibnamefont {Feldman}},\ }\href {\doibase
  10.5281/zenodo.10823282} {\enquote {\bibinfo {title} {Dataset for '{Mapping}
  twist-tuned multi-band topology in bilayer {WSe}\$\_2\$'},}\ } (\bibinfo
  {year} {2024}{\natexlab{b}})\BibitemShut {NoStop}%
\bibitem [{\citenamefont {Luryi}(1988)}]{luryiQuantumCapacitanceDevices1988}%
  \BibitemOpen
  \bibfield  {author} {\bibinfo {author} {\bibfnamefont {S.}~\bibnamefont
  {Luryi}},\ }\href {\doibase 10.1063/1.99649} {\bibfield  {journal} {\bibinfo
  {journal} {Applied Physics Letters}\ }\textbf {\bibinfo {volume} {52}},\
  \bibinfo {pages} {501} (\bibinfo {year} {1988})}\BibitemShut {NoStop}%
\bibitem [{\citenamefont {Jung}\ \emph {et~al.}(2017)\citenamefont {Jung},
  \citenamefont {Myoung}, \citenamefont {Park}, \citenamefont {Jeong},
  \citenamefont {Kim}, \citenamefont {Watanabe}, \citenamefont {Taniguchi},
  \citenamefont {Ha}, \citenamefont {Hwang},\ and\ \citenamefont
  {Park}}]{jungDirectProbingElectronic2017}%
  \BibitemOpen
  \bibfield  {author} {\bibinfo {author} {\bibfnamefont {S.}~\bibnamefont
  {Jung}}, \bibinfo {author} {\bibfnamefont {N.}~\bibnamefont {Myoung}},
  \bibinfo {author} {\bibfnamefont {J.}~\bibnamefont {Park}}, \bibinfo {author}
  {\bibfnamefont {T.~Y.}\ \bibnamefont {Jeong}}, \bibinfo {author}
  {\bibfnamefont {H.}~\bibnamefont {Kim}}, \bibinfo {author} {\bibfnamefont
  {K.}~\bibnamefont {Watanabe}}, \bibinfo {author} {\bibfnamefont
  {T.}~\bibnamefont {Taniguchi}}, \bibinfo {author} {\bibfnamefont {D.~H.}\
  \bibnamefont {Ha}}, \bibinfo {author} {\bibfnamefont {C.}~\bibnamefont
  {Hwang}}, \ and\ \bibinfo {author} {\bibfnamefont {H.~C.}\ \bibnamefont
  {Park}},\ }\href {\doibase 10.1021/acs.nanolett.6b03821} {\bibfield
  {journal} {\bibinfo  {journal} {Nano Letters}\ }\textbf {\bibinfo {volume}
  {17}},\ \bibinfo {pages} {206} (\bibinfo {year} {2017})}\BibitemShut
  {NoStop}%
\bibitem [{\citenamefont {Wang}\ \emph {et~al.}(2013)\citenamefont {Wang},
  \citenamefont {Meric}, \citenamefont {Huang}, \citenamefont {Gao},
  \citenamefont {Gao}, \citenamefont {Tran}, \citenamefont {Taniguchi},
  \citenamefont {Watanabe}, \citenamefont {Campos}, \citenamefont {Muller},
  \citenamefont {Guo}, \citenamefont {Kim}, \citenamefont {Hone}, \citenamefont
  {Shepard},\ and\ \citenamefont
  {Dean}}]{wangOneDimensionalElectricalContact2013}%
  \BibitemOpen
  \bibfield  {author} {\bibinfo {author} {\bibfnamefont {L.}~\bibnamefont
  {Wang}}, \bibinfo {author} {\bibfnamefont {I.}~\bibnamefont {Meric}},
  \bibinfo {author} {\bibfnamefont {P.~Y.}\ \bibnamefont {Huang}}, \bibinfo
  {author} {\bibfnamefont {Q.}~\bibnamefont {Gao}}, \bibinfo {author}
  {\bibfnamefont {Y.}~\bibnamefont {Gao}}, \bibinfo {author} {\bibfnamefont
  {H.}~\bibnamefont {Tran}}, \bibinfo {author} {\bibfnamefont {T.}~\bibnamefont
  {Taniguchi}}, \bibinfo {author} {\bibfnamefont {K.}~\bibnamefont {Watanabe}},
  \bibinfo {author} {\bibfnamefont {L.~M.}\ \bibnamefont {Campos}}, \bibinfo
  {author} {\bibfnamefont {D.~A.}\ \bibnamefont {Muller}}, \bibinfo {author}
  {\bibfnamefont {J.}~\bibnamefont {Guo}}, \bibinfo {author} {\bibfnamefont
  {P.}~\bibnamefont {Kim}}, \bibinfo {author} {\bibfnamefont {J.}~\bibnamefont
  {Hone}}, \bibinfo {author} {\bibfnamefont {K.~L.}\ \bibnamefont {Shepard}}, \
  and\ \bibinfo {author} {\bibfnamefont {C.~R.}\ \bibnamefont {Dean}},\ }\href
  {\doibase 10.1126/science.1244358} {\bibfield  {journal} {\bibinfo  {journal}
  {Science}\ }\textbf {\bibinfo {volume} {342}},\ \bibinfo {pages} {614}
  (\bibinfo {year} {2013})}\BibitemShut {NoStop}%
\bibitem [{\citenamefont {Lau}\ \emph {et~al.}(2022)\citenamefont {Lau},
  \citenamefont {Bockrath}, \citenamefont {Mak},\ and\ \citenamefont
  {Zhang}}]{lauReproducibilityFabricationPhysics2022}%
  \BibitemOpen
  \bibfield  {author} {\bibinfo {author} {\bibfnamefont {C.~N.}\ \bibnamefont
  {Lau}}, \bibinfo {author} {\bibfnamefont {M.~W.}\ \bibnamefont {Bockrath}},
  \bibinfo {author} {\bibfnamefont {K.~F.}\ \bibnamefont {Mak}}, \ and\
  \bibinfo {author} {\bibfnamefont {F.}~\bibnamefont {Zhang}},\ }\href
  {\doibase 10.1038/s41586-021-04173-z} {\bibfield  {journal} {\bibinfo
  {journal} {Nature}\ }\textbf {\bibinfo {volume} {602}},\ \bibinfo {pages}
  {41} (\bibinfo {year} {2022})}\BibitemShut {NoStop}%
\bibitem [{\citenamefont {Laturia}\ \emph {et~al.}(2018)\citenamefont
  {Laturia}, \citenamefont {Van~de Put},\ and\ \citenamefont
  {Vandenberghe}}]{laturiaDielectricPropertiesHexagonal2018}%
  \BibitemOpen
  \bibfield  {author} {\bibinfo {author} {\bibfnamefont {A.}~\bibnamefont
  {Laturia}}, \bibinfo {author} {\bibfnamefont {M.~L.}\ \bibnamefont {Van~de
  Put}}, \ and\ \bibinfo {author} {\bibfnamefont {W.~G.}\ \bibnamefont
  {Vandenberghe}},\ }\href {\doibase 10.1038/s41699-018-0050-x} {\bibfield
  {journal} {\bibinfo  {journal} {npj 2D Materials and Applications}\ }\textbf
  {\bibinfo {volume} {2}},\ \bibinfo {pages} {6} (\bibinfo {year}
  {2018})}\BibitemShut {NoStop}%
\bibitem [{\citenamefont {Pierret}\ \emph {et~al.}(2022)\citenamefont
  {Pierret}, \citenamefont {Mele}, \citenamefont {Graef}, \citenamefont
  {Palomo}, \citenamefont {Taniguchi}, \citenamefont {Watanabe}, \citenamefont
  {Li}, \citenamefont {Toury}, \citenamefont {Journet}, \citenamefont {Steyer},
  \citenamefont {Garnier}, \citenamefont {Loiseau}, \citenamefont {Berroir},
  \citenamefont {Bocquillon}, \citenamefont {Fève}, \citenamefont {Voisin},
  \citenamefont {Baudin}, \citenamefont {Rosticher},\ and\ \citenamefont
  {Plaçais}}]{pierretDielectricPermittivityConductivity2022}%
  \BibitemOpen
  \bibfield  {author} {\bibinfo {author} {\bibfnamefont {A.}~\bibnamefont
  {Pierret}}, \bibinfo {author} {\bibfnamefont {D.}~\bibnamefont {Mele}},
  \bibinfo {author} {\bibfnamefont {H.}~\bibnamefont {Graef}}, \bibinfo
  {author} {\bibfnamefont {J.}~\bibnamefont {Palomo}}, \bibinfo {author}
  {\bibfnamefont {T.}~\bibnamefont {Taniguchi}}, \bibinfo {author}
  {\bibfnamefont {K.}~\bibnamefont {Watanabe}}, \bibinfo {author}
  {\bibfnamefont {Y.}~\bibnamefont {Li}}, \bibinfo {author} {\bibfnamefont
  {B.}~\bibnamefont {Toury}}, \bibinfo {author} {\bibfnamefont
  {C.}~\bibnamefont {Journet}}, \bibinfo {author} {\bibfnamefont
  {P.}~\bibnamefont {Steyer}}, \bibinfo {author} {\bibfnamefont
  {V.}~\bibnamefont {Garnier}}, \bibinfo {author} {\bibfnamefont
  {A.}~\bibnamefont {Loiseau}}, \bibinfo {author} {\bibfnamefont {J.-M.}\
  \bibnamefont {Berroir}}, \bibinfo {author} {\bibfnamefont {E.}~\bibnamefont
  {Bocquillon}}, \bibinfo {author} {\bibfnamefont {G.}~\bibnamefont {Fève}},
  \bibinfo {author} {\bibfnamefont {C.}~\bibnamefont {Voisin}}, \bibinfo
  {author} {\bibfnamefont {E.}~\bibnamefont {Baudin}}, \bibinfo {author}
  {\bibfnamefont {M.}~\bibnamefont {Rosticher}}, \ and\ \bibinfo {author}
  {\bibfnamefont {B.}~\bibnamefont {Plaçais}},\ }\href {\doibase
  10.1088/2053-1591/ac4fe1} {\bibfield  {journal} {\bibinfo  {journal}
  {Materials Research Express}\ }\textbf {\bibinfo {volume} {9}},\ \bibinfo
  {pages} {065901} (\bibinfo {year} {2022})}\BibitemShut {NoStop}%
\end{thebibliography}%

\end{document}